\documentclass[twocolumn]{aastex631}

\usepackage{amsmath}
\usepackage{enumitem}
\usepackage{subfigure}

\submitjournal{\apj}

\usepackage{soul}

\shorttitle{CCSN simulations with reduced nucleosynthesis networks}
\shortauthors{Nav\'o, Reichert, Obergaulinger, \& Arcones}

\begin{document}
\title{Core-collapse Supernova Simulations with Reduced Nucleosynthesis Networks}

\author[0000-0002-8984-0211]{Gerard Nav\'o}
\affiliation{Institut f{\"u}r Kernphysik, Technische Universit{\"a}t Darmstadt, Schlossgartenstr. 2, Darmstadt 64289, Germany}
\correspondingauthor{Gerard Nav\'o}
\email{gnavo@theorie.ikp.physik.tu-darmstadt.de}
\author[0000-0001-6653-7538]{Moritz Reichert}
\affiliation{Departament d'Astronomia i Astrof{\'i}sica, Universitat de Val{\`e}ncia, \\
 Edifici d{\'{}}Investigaci{\'o} Jeroni Munyoz, C/ Dr.~Moliner, 50,
 E-46100 Burjassot (Val{\`e}ncia), Spain}
 
\author[0000-0001-5664-1382]{Martin Obergaulinger}
\affiliation{Departament d'Astronomia i Astrof{\'i}sica, Universitat de Val{\`e}ncia, \\
 Edifici d{\'{}}Investigaci{\'o} Jeroni Munyoz, C/ Dr.~Moliner, 50,
 E-46100 Burjassot (Val{\`e}ncia), Spain}
 
\author[0000-0002-6995-3032]{Almudena Arcones}
\affiliation{Institut
f{\"u}r Kernphysik, Technische Universit{\"a}t Darmstadt, Schlossgartenstr. 2, Darmstadt 64289, Germany}
\affiliation{GSI Helmholtzzentrum f{\"u}r Schwerionenforschung GmbH, Planckstr. 1, Darmstadt 64291, Germany}



\begin{abstract}

 We present core-collapse supernova simulations including nuclear reaction networks that impact explosion dynamics and nucleosynthesis. The different composition treatment can lead to changes in the neutrino heating in the vicinity of the shock by modifying the number of nucleons and thus the neutrino-opacity of the region. This reduces the ram pressure outside the shock and allows an easier expansion. The energy released by the nuclear reactions during collapse also slows down the accretion and aids the shock expansion. In addition, nuclear energy generation in the postshocked matter produces up to $20\%$ more energetic explosions. Nucleosynthesis is affected due to the different dynamic evolution of the explosion. Our results indicate that the energy generation from nuclear reactions helps to sustain late outflows from the vicinity of the proto-neutron star, synthesizing more neutron-rich species. Furthermore, we show that there are systematic discrepancies between the ejecta calculated with in-situ and ex-situ reaction networks. These differences stem from the intrinsic characteristics of evolving the composition in hydrodynamic simulations or calculating it with Lagrangian tracer particles. The mass fractions of some Ca, Ti, Cr, and Fe isotopes are consistently underproduced in postprocessing calculations, leading to different nucleosynthesis paths. Our results suggest that large in-situ nuclear reaction networks are important for a realistic feedback of the energy generation, the neutrino heating, and a more accurate ejecta composition.
 
\end{abstract}




\section{Introduction} \label{sec:intro}
Massive stars ($M\gtrsim 8 \,\mathrm{M_\odot}$) usually end their life in core-collapse supernovae (CCSN). These explosive events that are triggered by the collapse of the iron core and release several solar masses of products that were synthesized during the stellar evolution and the explosion itself into the interstellar medium. Thus, they play a critical role in the chemical history of the universe and have been a matter of study for many decades. In order to investigate the mechanisms involved, radiation-hydrodynamic simulations have been performed \citep[for extended reviews, see, e.g.,][]{Bethe90,Kotake_2006,Janka_12_review,mueller_20_review,Burrows_21_review}, following different evolutionary stages from collapse to shock breakout. Nevertheless, the whole puzzle is still far from being completed.

In recent years, significant advances have been reported in multidimensional simulations \citep[e.g.,][]{lentz15,janka16,roberts_3D,oconnor_couch_3D,martin_aloy_magnetorotational,burrows_3d,takami_3d,sandoval_21,nakamura_22}, impact of magnetic fields \citep[e.g.,][]{moesta_15,martin_just_aloy,takami_3d,varma22}, presupernova models \citep[e.g.,][]{muellerB_2017,fields21,yoshida_progenitor_convection,vartanyan_progenitorsCCSN}, neutrino transport \citep[for an extended review, see][]{mezzacapa_neutrino_transport_review} and neutrino reactions \citep[e.g.,][and references therein]{Balasi_2015}, high-density equations of state (EOS) \citep[e.g.,][]{Schneider_2019,hanna_sabrina} or nucleosynthesis calculations \citep[e.g.,][]{Eichler_2017,wanajo18,curtis_19,sieverding20,witt_paper21,moritz_magneto}. In this work, we focus on the impact of the different ways to treat nuclear reactions and the associated energy generation in the simulations.

A nuclear reaction network evolves the abundances of a set of $N$ species by a system of $N$ coupled ordinary differential equations including all the reactions between the species. Therefore, its size depends on the nature of the environment. In CCSNe, it can include some several hundred nuclei \citep[e.g.,][]{woosley_95,Rauscher_2002}. Thus, including nuclear reaction networks in multidimensional CCSN simulations is very challenging because of the computational cost of evolving them together with the hydrodynamics. Therefore, hydrodynamic simulations usually employ a simple treatment of the composition. A detailed nucleosynthesis can be obtained by postprocessing \citep[e.g.,][]{winteler_winnet,lippuner17a, reichert_inprep} using Lagrangian tracer particles. 

If the temperatures are high enough ($ T \gtrsim 5-6 \, \mathrm{GK}$), fusion reactions and photodisintegrations are in so-called nuclear statistical equilibrium (NSE). Within this equilibrium, the production rate of a specific nuclear species equals its destruction rate, and for constant thermodynamic conditions, the composition is also constant over time (neglecting weak reactions). In NSE, the abundances are direct functions of the density, temperature, and electron fraction ($Y_\mathrm{e}$) conditions. Thus, instead of the costly evolution of the reaction network, they can be used at high temperatures and be included in high-density EOS tables. If the temperature is lower, NSE no longer holds, and other methods are necessary to estimate the composition and energy generation. In this case, other simplified treatments are used. The so-called flashing scheme \citep{flashing-scheme} takes into account neutrons, protons, $\mathrm{\alpha}$-particles, and a characteristic nucleus that depends on the thermodynamic conditions, considering instantaneous burning. For example, when silicon burning occurs, the silicon mass fraction ($X(^{28}\mathrm{Si})$) is immediately transformed into $^{56}\mathrm{Ni}$ and set to $X(^{28}\mathrm{Si})=0$. While the composition assumed by the flashing scheme is only a rough approximation, this scheme implicitly accounts for the generation or consumption of nuclear energy by the instantaneous burning between the nuclei that is assumed to happen at the threshold temperature. Thus, explicit source terms for nuclear energy generation are not required. An alternative is to use reduced reaction networks \citep[e.g.,][]{Mueller86,Hix-thielemann1999,timmes2000} in order to track a small set of nuclei together with the hydrodynamics. The considered nuclei are chosen to involve the main reactions that release or consume internal energy and to represent the main contributions to the baryonic part of the pressure and neutrino opacities \citep{opacities-cerno}. In practice, $\alpha$-chains are the most commonly used for this purpose.

Several groups have employed reduced networks in their works \citep[e.g.,][]{bruenn16,bruenn20,nakamura2014,Couch15,harris17,Wongwathanarat17,sandoval_21}. \citet{bruenn2006} mentioned that when including a $14$ $\mathrm{\alpha}$-chain in the hydrodynamics, nuclear burning in the oxygen layer deposited additional pressure in the vicinity of the shock and assisted its expansion. \citet{nakamura2014} performed simulations with a simplified light-bulb neutrino treatment using a $13$ isotope $\mathrm{\alpha}$-network to study how the energy released by nuclear reactions affects the dynamics of the explosion. They concluded that energy produced by the infalling material behind the shock could aid the explosion, especially in models with marginal explosions. A large network was included for the first time by \citet{harris17} in 2D simulations with accurate $\mathrm{\nu}$-transport. They used a $14$ $\mathrm{\alpha}$-chain and a $150$ isotope network in axisymmetric (2D) models and studied the uncertainties derived from postprocessing nucleosynthesis. Their results showed the limitations of using postprocessing Lagrangian tracer particles and support including reaction networks in the simulations.

In this work, the aforementioned studies motivated us to investigate in detail, in a state-of-the-art CCSN code, how the different treatments of the composition employed in the simulations at low temperatures affect the dynamics of the explosion and the nucleosynthesis outcomes. In Section~\ref{sec:methods} we introduce the radiation-hydrodynamics code we used for the simulations (Section~\ref{subsec: HD}), the reduced network module implementation and the chosen networks (Section~\ref{subsec: red net implementation}), and the different models (Section~\ref{subsec: models}). In Section~\ref{sec:results} we discuss the results, and in Section~\ref{sec: conclusions} we close with our conclusions and final remarks.

\section{Methods}
\label{sec:methods}
\subsection{Hydrodynamic code}
\label{subsec: HD}
The simulations are performed using the code \textsc{Aenus-Alcar} \citep{Aenus-Alcar-Just,martin350c}, which combines special relativistic magnetohydrodynamics (MHD), a pseudo-relativistic gravitational potential, and state-of-the-art two-moment (M1) neutrino transport. 

Prior to the implementation of the reduced network module, \textsc{Aenus-Alcar} employed a nuclear EOS for matter at high densities ($\rho > \rho_\mathrm{th} \sim 10^{7-8} \mathrm{g\,cm}^{-3}$) assuming NSE between protons, neutrons, alpha particles, and a representative heavy nucleus. At lower densities,  ($\rho \le \rho_\mathrm{th}$), a Helmholtz-type EOS was used \citep[see, e.g.,][]{Timmes_eos1,Timmes_eos_2000}, which takes into account leptonic, photonic, and baryonic contributions. For the latter, a version of the flashing scheme \citep{flashing-scheme} calculated the composition of the neutrons, protons, and a characteristic nucleus that is $^{28}\mathrm{Si}$ or $^{56}\mathrm{Ni}$, depending on the thermodynamic conditions.  In this work, we have added a nuclear reaction network module to describe the composition outside of the NSE regime more accurately.

\subsection{Reduced Network Implementation} 
\label{subsec: red net implementation}

\subsubsection{\textsc{ReNet}}

\textsc{ReNet} is a highly flexible nuclear reaction network code that allows to calculate abundance flows for networks of different sizes and complexities. Similar to previous works in the literature \citep{Mueller86,Hix-thielemann1999,timmes2000}, it solves the set of ordinary first-order differential equations in an implicit way. The integration scheme in \textsc{ReNet} is a first-order implicit Euler scheme \citep[as described in e.g.,][]{Hix-thielemann1999,lippuner17a}. The convergence criterion of the necessary Newton-Raphson scheme is given by the mass conservation, \mbox{$\sum Y_i A_i = \sum X_i =1$}, with the abundance $Y_i$ and mass number $A_i$. This is identical to what is used in \citet{lippuner17a}. In contrast to postprocessing reaction networks, the thermodynamic quantities in the next time step are unknown for in-situ reaction networks. Therefore, \textsc{ReNet} burns hydrostatically in each time step with the temperature and density of the current step. The computational cost of nuclear networks can be reduced by taking advantage of sparse matrices \citep[e.g.,][]{Hix-thielemann1999,lippuner17a}. However, for the reduced reaction networks used here, the computational overhead for this is large, and we consequently use solvers for dense matrices. Within \textsc{ReNet} we furthermore assume some species to be in steady state in order to consider them indirectly and to avoid using a differential equation for them. This allows us to save computational cost \citep[for more information, see Appendix~\ref{sec: appendix network} and see,][for a similar approach]{approx19,Timmes1999,Paxton2011}. 

Additionally, we compute the energy released by the nuclear reactions through
\begin{equation}
\label{eq: Enuc}
    \dot{E}_\text{nuc}=-\sum_i N_\mathrm{A} {\Delta m}_i c^2 \dot{Y}_i,
\end{equation} 
where $\dot{E}_\text{nuc}$ is the variation in the internal energy due to the nuclear reactions, ${\Delta m}_i$ is the mass excess of each species, and $\dot{Y}_i$ is the abundance time variation of the nucleus $i$.

\begin{figure}[tb]
    \centering
    \includegraphics[width=1.\columnwidth]{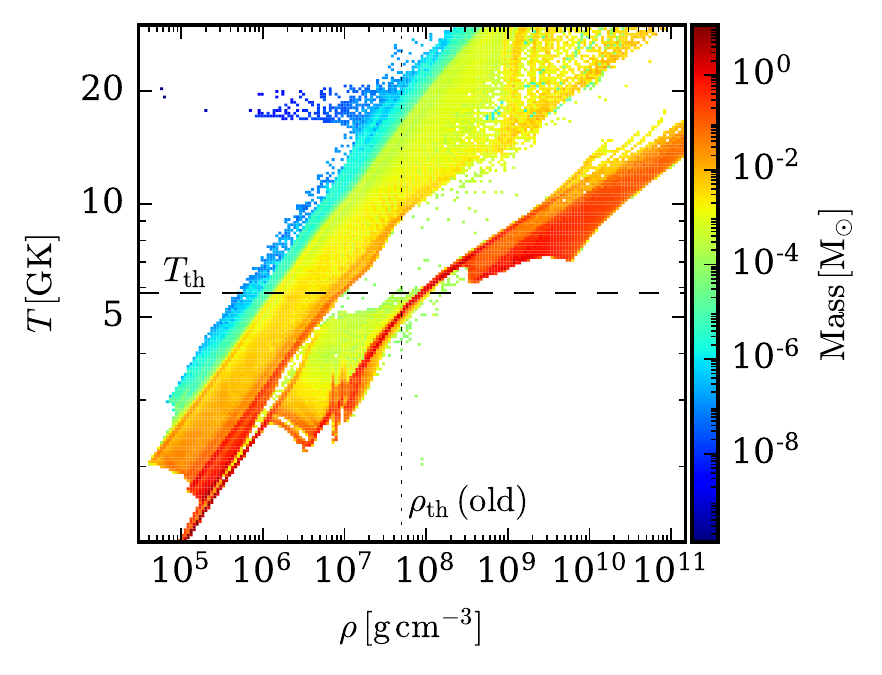}
    \caption{Density and temperature achieved in a characteristic CCSN simulation. It focuses on the region of the $\rho$-$T$ plane close to the transition between the two EOS. The color depicts the amount of mass that fulfilled each $(\rho,T)$ condition in the run. The dashed line indicates the temperature threshold criterion introduced in this work. Finally, the dotted line corresponds to the density threshold criterion present in previous versions of \textsc{Aenus-Alcar}.}
    \label{fig:transition}
\end{figure}
\subsubsection{\textsc{ReNet} in \textsc{Aenus-Alcar}}
\label{subsubsec: Aenus Implementation}

As mentioned above, previous versions of \textsc{Aenus-Alcar} distinguish between two different density ranges where two different EOS regimes are used, one of which assumes NSE. In contrast, for the composition, the switch between NSE and the non-NSE regime using the reaction network is based on temperature rather than density.  We found numerical artifacts in regions where the two criteria contradicted each other, e.g., zones of low density but high temperature for which the EOS would not select NSE, but the composition would be given by NSE instead of the network.  
This is the case for a large number of  zones in the top left part of the phase diagram shown for a typical CCSN simulation in Figure~\ref{fig:transition}.
To avoid this an inconsistency, we use the same temperature criteria in the NSE-network and the EOS transitions, such that above a certain temperature threshold ($T_{\mathrm{th}}$), $T>T_{\mathrm{th}} \sim 5-6 \, \mathrm{GK}$, NSE is assumed both for the composition and the EOS.  Otherwise,   the network is employed together with the subnuclear EOS.  Although in principle, the latter case would also apply to very cold zones with densities up to and beyond the nuclear saturation density, in practice, this regime (bottom right part of the figure) is not achieved in CCSN and thus is of no concern.

\begin{figure}[t]
    \centering
    \includegraphics[width=1.\columnwidth]{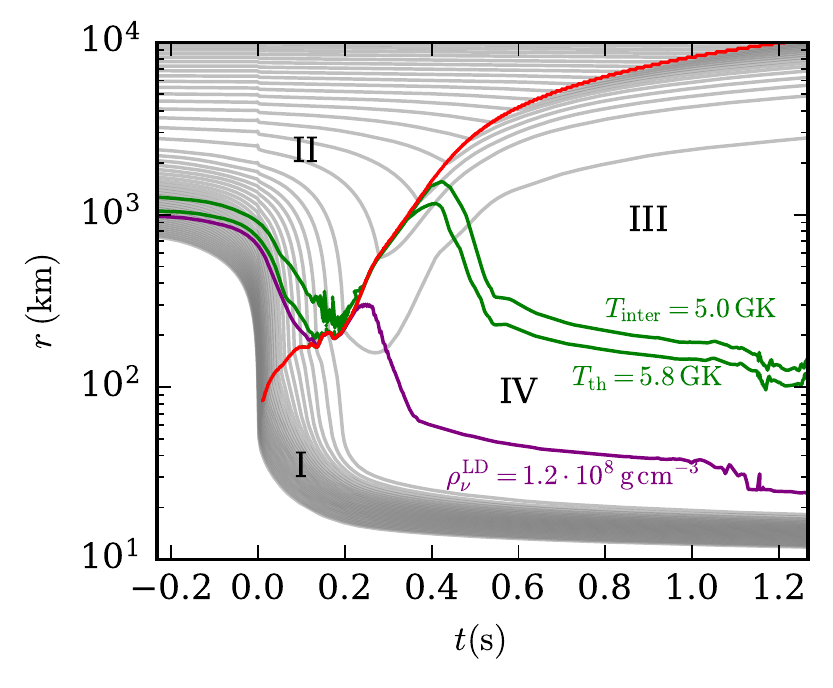}
    \caption{Schematic representation of the different regions of interest in this work. It shows the mass shell evolution (gray) of a characteristic 1D model. The red line shows the shock evolution. The temperatures involved in the transition between the NSE and the network regimes are depicted in green. The purple line shows the evolution of $\rho_\nu^\mathrm{LD}$, the density at which we switch off the neutrino absorption in the toy models described in Section~\ref{subsec: models}.}
    \label{fig:sketch}
\end{figure}
In order to prevent discontinuities in the thermodynamic quantities such as the pressure or the internal energy, we interpolate them linearly between the two EOSs in the temperature regime $T_{\mathrm{th}} > T > T_{\mathrm{inter}}$ , where $T_{\mathrm{inter}}$ is defined as the lower-limit temperature for the interpolation between EOSs (for a schematic representation, see Figure~\ref{fig:sketch}). 
Furthermore, at $T=T_{\mathrm{th}}$, the initial values for the network need to be given. These seeds are provided by an NSE solver that considers the same nuclei as the nuclear reaction network. When the temperature in a grid cell drops below $T_{\mathrm{th}}$, the network module evolves the composition further until the end of the simulation or until the conditions for NSE are reached again.

The energy from the nuclear reactions (Eq.(\ref{eq: Enuc})) is coupled as a source term to the hydrodynamics energy equation. In order to avoid thermodynamic inconsistencies, it is done for temperatures $T \le T_{\mathrm{inter}}$. We consider two different ways to include it. On the one hand, the feedback of the nuclear energy from the reaction network is considered at every cell with $T \le T_{\mathrm{inter}}$ (regions II and III in Figure~\ref{fig:sketch}). On the other hand, the nuclear energy generation is taken into account only after the shock has passed the cell, i.e., in the postshocked gas (region III). For this purpose, we use a simple shock-detection algorithm. We additionally consider the case without feedback of the nuclear energy generation from the nuclear reaction network.

\begin{deluxetable}{ccccccc}[tb!]
\tablecaption{List of RN16 nuclei \label{Tab:RN16 species}}
\tablehead{\multicolumn{7}{c}{RN16 Network}}
\startdata
$\mathrm{n}$ & $^{1}\mathrm{H}$ & $^{4}\mathrm{He}$ & $^{12}\mathrm{C}$ & $^{16}\mathrm{O}$ & $^{20}\mathrm{Ne}$ & $^{24}\mathrm{Mg}$ \\ 
$^{27}\mathrm{Al}^{\dagger}$ & $^{28}\mathrm{Si}$ & $^{31}\mathrm{P}^\dagger$ &  $^{32}\mathrm{S}$ & $^{35}\mathrm{Cl}^\dagger$ & $^{36}\mathrm{Ar}$ & $^{39}\mathrm{K}^\dagger$ \\ 
$^{40}\mathrm{Ca}$ & $^{43}\mathrm{Sc}^\dagger$ & $^{44}\mathrm{Ti}$ & $^{47}\mathrm{V}^\dagger$ & $^{48}\mathrm{Cr}$ & $^{51}\mathrm{Mn}^\dagger$ & $^{52}\mathrm{Fe}$ \\ 
$^{53}\mathrm{Fe}^\dagger$ & $^{54}\mathrm{Fe}$ & $^{55}\mathrm{Co}^\dagger$ & $^{56}\mathrm{Ni}$\\
\enddata
\tablecomments{Steady-state nuclei are marked with a dagger.}
\end{deluxetable}
\begin{deluxetable}{rrrr}[tb!]
\tablecaption{List of RN94 nuclei \label{Tab:RN94 species}}
\tablehead{\multicolumn{4}{c}{RN94 Network}}
\startdata
$\mathrm{n}$ & $^{1}\mathrm{H}$ & $^{4}\mathrm{He}$ & $^{12}\mathrm{C}$ \\
$^{16}\mathrm{O}$ & $^{20}\mathrm{Ne}$ & $^{23}\mathrm{Mg}^\dagger$ & $^{24}\mathrm{Mg}$ \\ 
$^{27}\mathrm{Al}^{\dagger}$ & $^{28,30}\mathrm{Si}$ & $^{27,29}\mathrm{Si}^\dagger$ & $^{29,31}\mathrm{P}^\dagger$ \\
$^{30-34}\mathrm{S}$ & $^{31,33}\mathrm{S}^\dagger$ & $^{33,35}\mathrm{Cl}^\dagger$ &
$^{34-38}\mathrm{Ar}$ \\
$^{35,37}\mathrm{Ar}^\dagger$ & $^{37,39}\mathrm{K}^\dagger$ & $^{38-42,48-52}\mathrm{Ca}$ & $^{39,41,49,51}\mathrm{Ca}^\dagger$ \\
$^{41,43,49-55}\mathrm{Sc}^\dagger$ & $^{42-56}\mathrm{Ti}$ & $^{43-55}\mathrm{Ti}^\dagger$ & $^{45-57}\mathrm{V}^\dagger$ \\
$^{46-58}\mathrm{Cr}$ & $^{47-57}\mathrm{Cr}^\dagger$ & $^{49-61}\mathrm{Mn}^\dagger$ & $^{50-64}\mathrm{Fe}$ \\
$^{51-63}\mathrm{Fe}^\dagger$ & $^{53-67}\mathrm{Co}^\dagger$ & $^{54-70}\mathrm{Ni}$ &  $^{55-69}\mathrm{Ni}^\dagger$ \\
$^{61-73}\mathrm{Cu}^\dagger$ & $^{60-74}\mathrm{Zn}$ & $^{59-73}\mathrm{Zn}^\dagger$ & $^{63-77}\mathrm{Ga}^\dagger$ \\
$^{64-80}\mathrm{Ge}$ & $^{65-79}\mathrm{Ge}^\dagger$ & $^{67-83}\mathrm{As}^\dagger$ 
& $^{68-84}\mathrm{Se}$  \\
$^{69-83}\mathrm{Se}^\dagger$  & $^{75-85}\mathrm{Br}^\dagger$  & $^{76-86}\mathrm{Kr}$ & $^{77-85}\mathrm{Kr}^\dagger$ \\
$^{79-87}\mathrm{Rb}^\dagger$ & $^{80-88}\mathrm{Sr}$ & $^{81-87}\mathrm{Sr}^\dagger$ & $^{85-89}\mathrm{Y}^\dagger$ \\
$^{86-90}\mathrm{Zr}$ & $^{87,89}\mathrm{Zr}^\dagger$ & $^{91}\mathrm{Nb}^\dagger$ & $^{92}\mathrm{Mo}$\\
\enddata
\tablecomments{Steady-state nuclei are marked with a dagger. We use the $^{i-j}X$ notation to account for the species in $\{{^iX},{^{i+2}X},...,{^jX}\}$.}
\end{deluxetable}

\subsubsection{Networks}
\label{subsec: networks}
In this work, we implemented two reaction network configurations into \textsc{Aenus-Alcar}: a $16\text{-}\alpha$ chain (RN16), and a 94-species network (RN94). RN16 is similar to the widely used $19$ isotope approximation network from \citet{approx19}, which approximately reproduces the nuclear energy generation within CCSN simulations and therefore provides feedback to the total energy of the system. RN94 is able to track the main species synthesized in the CCSNe \citep[see][]{Eichler_2017,harris17}, with $Y_e$ between $0.4$ and $0.6$. Hence, it allows us to track slightly neutron- and proton- rich trajectories along stability up to $^{92}\mathrm{Mo}$. In addition, $148$ species are considered in steady state, reproducing a network consisting of $242$ nuclei. Therefore, at the time of writing, we present the most complete network evolved in CCSN simulations with state-of-the-art M1 neutrino transport. The species included in both networks are listed in Table~\ref{Tab:RN16 species} and Table~\ref{Tab:RN94 species}.

For comparison, the nucleosynthesis is additionally computed in postprocessing with the full nuclear reaction network \textsc{WinNet} \citep{reichert_inprep} and the reduced networks RN16 and RN94. We consider the unbound matter, i.e., matter with positive total energy and positive radial velocity, at the final time of the simulation ($t_\mathrm{f}=1.5\,\mathrm{s}$). The evolution of the ejecta is followed by Lagrangian tracer particles calculated backward in time, as described in \citet{reichert22} (see also \citealt{Sieverding2023} for an in-depth analysis of the uncertainties that can arise with this method). At the final simulation time, we place them at random positions in all cells flagged as unbound. The total mass contained in a cell is distributed equally among its tracers. Their number is set such that they have a maximum mass, $M=10^{-4} \, \mathrm{M_\odot}$, and each cell contains at least four tracers. At the end, the nucleosynthesis of each tracer particle is weighted with its corresponding mass.

\begin{deluxetable}{lcccc}[tb!]
\tablecaption{List of the different models \label{Tab:models}}
\tablehead{
\colhead{Model} & \colhead{Comp} & \colhead{$\dot{E}_\text{nuc}$} & \colhead{$Q^\mathrm{LD}_\mathrm{\nu}$} & \colhead{Dim}}
\startdata
$\mathrm{1D\_flsh\_noQ_\nu^{LD}}$  & $\mathrm{flashing}^\dagger$ & no & no  & 1D \\
$\mathrm{1D\_RN16\_noQ_\nu^{LD}}$  & RN16 & no & no & 1D \\
$\mathrm{1D\_RN94\_noQ_\nu^{LD}}$  & RN94 & no & no & 1D \\
$\mathrm{1D\_SFHo\_noQ_\nu^{LD}} $ & NSE & no & no& 1D \\
1D\_flsh  & $\mathrm{flashing}^\dagger$ & no & yes & 1D \\
1D\_RN16  & RN16 & no & yes & 1D \\
1D\_RN94  & RN94 & no & yes & 1D \\
1D\_SFHo  & NSE & no &yes& 1D \\
1D\_RN16E  & RN16 & yes &yes& 1D \\
1D\_RN16e  & RN16 & p.s & yes & 1D \\
1D\_RN94E  & RN94 & yes & yes & 1D \\
1D\_RN94e  & RN94 & p.s &yes & 1D \\
1D\_flshE & flashing & yes & yes  & 1D \\
2D\_flsh  & $\mathrm{flashing}^\dagger$ & no & yes &  2D \\
2D\_RN16E &  RN16 & yes & yes& 2D \\
2D\_RN94E  & RN94 & yes & yes & 2D \\
\enddata
\tablecomments{The second column shows the treatment of the composition. The third column specifies whether the energy generation from the nuclear reactions is taken into account. p.s indicates that it is switched on only in the postshocked region. The fourth column indicates whether the neutrino interactions are taken into account at $\rho<1.2\cdot10^8\, \mathrm{g\,cm^{-3}}$. Finally, the last column states for the dimensionality of the simulations.}
\end{deluxetable}

\subsection{Models}
\label{subsec: models}

We have performed simulations ($t_\mathrm{f}=1.5\, \mathrm{s}$) using the solar metallicity $20 \, \mathrm{M_\odot}$ mass progenitor from \citet{WH2007}. We map its precollapse composition to those of our two networks. We consider the EOS transition at $T_{\mathrm{th}}=5.8 \, \mathrm{GK}$\footnote{We tested several transition temperatures. A higher transition temperature ( $T_\mathrm{th}=6.5 \, \mathrm{GK}$) does not change the dynamics significantly. When decreasing it ($T_\mathrm{th}=5 \, \mathrm{GK}$), the alpha-rich freeze out is not well characterized.}, given that NSE breaks down around this temperature, and $T_{\mathrm{inter}}=5.0 \, \mathrm{GK}$. Since this work is focused on the impact of the composition treatment at $T < T_{\mathrm{th}}$, all of the models have the same configuration at high temperatures (areas I and IV in Figure~\ref{fig:sketch}). We apply the SFHo EOS \citep{sfho} and assume the composition provided by it in NSE. We consider the main neutrino-matter interactions \citep[][]{Aenus-Alcar-Just,Just__2018__MonthlyNoticesoftheRoyalAstronomicalSociety__CoreCollapseSupernovaSimulationsinOneandTwoDimensionsComparisonofCodesandApproximations} contributing to the neutrino energy deposition, $Q_{\mathrm{\nu}}$, which is critical for triggering the explosion. It is well known that one-dimensional (1D) spherically symmetric simulations do not explode due to the lack of convection and hydrodynamical instabilities \citep[e.g.,][]{Janka_12_review}.  Therefore, we add additional heating factor, $\mathit{HF}=2.8$, in the gain region to launch explosions in all of the 1D \citep[see, e.g.,][for a similar approach]{MaxW_thesis}: $Q_\mathrm{\nu}^{\mathrm{gain,}\mathit{HF}}= \mathit{HF}\cdot Q_\mathrm{\nu}^\mathrm{gain}$, where $Q_\mathrm{\nu}^\mathrm{gain}$ is the neutrino energy deposition in the gain layer. In 2D models, we set $\mathit{HF}=1$ and do not add any additional neutrino heating.

In order to understand the impact of nuclear reaction networks coupled to the hydrodynamics, we have varied the treatment of the composition and nuclear energy generation at $T \le T_{\mathrm{th}}$. Four approaches have been used to describe the composition at low temperatures and densities, where a network is necessary:
\begin{itemize}
    \item The crudest description is to keep using the high-density (or nuclear) EOS table at low temperatures as well. This is commonly done in many supernova simulations that focus on the early explosion phase \citep[e.g.,][]{oconnor_couch_3D}. The composition in EOS tables typically consists of neutrons, protons, alpha particles, and a representative nucleus. This nucleus is obtained from assuming NSE or from single nucleus approximation \citep[SNA; see][for a discussion of the impact of these two treatments]{schneider17}. These models are marked ``SFHo'' and their composition corresponds to NSE.
    \item The flashing scheme was the improved composition treatment that has been used in the previous studies with ALCAR. In this set of models, the version of the flashing scheme employed does not include the nuclear energy implicitly for comparison purposes. Therefore, to  distinguish it from the original version \citep{flashing-scheme}, we refer to it as $\mathrm{flashing}^\dagger$, and the models are labeled \_flsh.
    \item We use the two reduced networks described in Section~\ref{subsec: networks}. The model names are RN16 and RN94.
\end{itemize}
The composition obtained with the four approaches is different and influences the final abundances, but also the dynamics. Variations in the abundances of nuclei heavier than alphas have a minor contribution to the pressure because it is dominated by radiation and electrons, and the ion contribution is small. However, the changes in the number of neutrons and protons can impact the energy deposited by neutrinos. Therefore, we also ran a few toy models without the neutrino absorption at low densities, $\rho<1.2\cdot10^8\, \mathrm{g\,cm^{-3}}$ (models with no $Q_\nu^\mathrm{LD}$), i.e, $Q_\nu=0$ in regions II, III, and IV of Figure~\ref{fig:sketch}.

Finally, we also used various models to explore the impact of the energy generation by nuclear reactions when using the two reduced networks in 1D and 2D.  We consider $\dot{E}_\text{nuc}$ in the two ways introduced in Section~\ref{subsubsec: Aenus Implementation}: everywhere with $T<T_\mathrm{inter}$ (regions II and III in Figure~\ref{fig:sketch}) where the composition is determined by the reduced networks (models labeled E) and only between $T<T_\mathrm{inter}$ and the shock, i.e., only region III (models labeled p.s.). We employ these two different $\dot{E}_\text{nuc}$ configurations in order to distinguish between the impact in the infalling layers and in the postshock region. In addition, we ran a model with the flashing scheme including its nuclear energy generation (1D\_flshE) for comparison.

We perform the 1D simulations using a grid with $n_\mathrm{r}=408$ zones that are logarithmically spaced in the radial direction with a central grid width of $\Delta r = 400 \, \mathrm{m}$ and a maximum radius of $R_{\mathrm{out}} \simeq 9.05 \times 10^5 \, \mathrm{km}$. The 2D simulations run on a grid of $n_r=400$ zones in the radial direction and $n_{\theta}=128$ in the angular direction. We consider the employed resolution sufficient to adequately resolve the hydrodynamic quantities \citep[e.g.,][]{martin350c}. 

The models computed are listed in Table~\ref{Tab:models}. The overview of the results is given in Figure~\ref{fig: Rshock_Eexpl} where we present the shock and explosion energy evolution for all models. In the following sections, we discuss the impact of the composition and nuclear reactions on the dynamics and final abundances.

\section{Results}
\label{sec:results}

\subsection{Impact on the dynamics}
\label{sct:imp_dynamics}
In this section, we study the effects that reduced networks have on the dynamics of the explosion. We show how the different treatments of the composition lead to changes in the neutrino absorption that modify the dynamics (Section~\ref{subsubsct: composition_dyn}) and the impact of the energy released by nuclear reactions on the dynamics of the explosion in 1D (Section~\ref{subsubsct: engen1_dyn}) and 2D (Section~\ref{subsubsct: engen2_dyn})
\subsubsection{Composition}
\label{subsubsct: composition_dyn}
First, we detail the impact that the different composition treatments have on the dynamics of the explosion in the models introduced in Section~\ref{subsec: models}.
\begin{figure*}[p]
    \centering
    \includegraphics[width=2.\columnwidth]{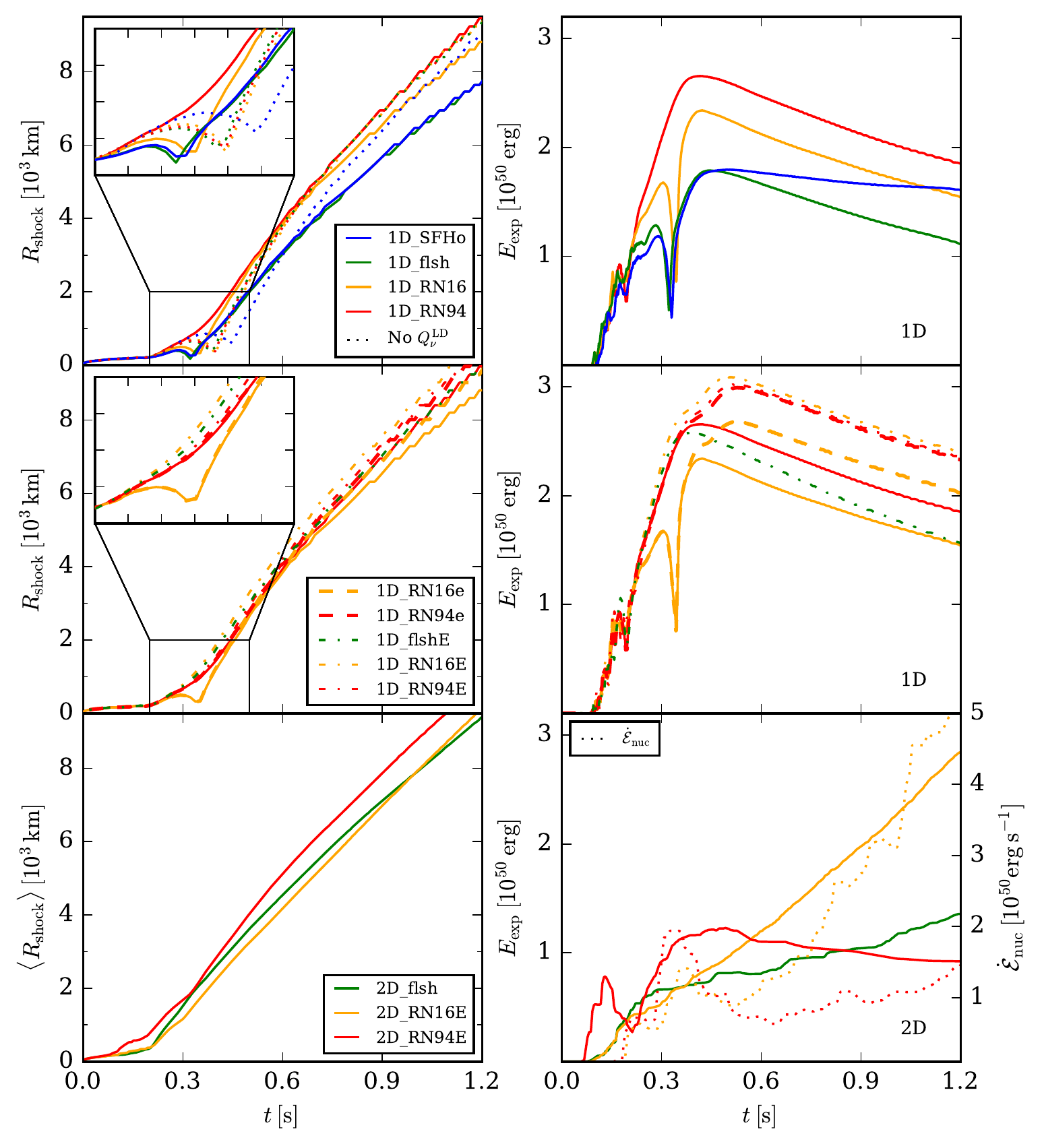}
    \caption{Shock evolution (left panels) and explosion energy (right panels) of the models discussed in this work, summarized in Table~\ref{Tab:models}. The different colors represent each treatment of the composition. The solid lines in the upper and middle panels stand for 1D models without $\dot{E}_\text{nuc}$, e.g. 1D\_RN16. The dotted lines correspond to models without $\dot{E}_\text{nuc}$ or $Q^{\mathrm{LD}}_{\mathrm{\nu}}$, e.g. $\mathrm{1D\_RN16\_noQ_\nu^{LD}}$. The dashed lines indicate the 1D\_RN16e and the 1D\_RN94e models. The dash-dotted lines correspond to the 1D\_RN16E, the 1D\_RN94E, and the 1D\_flshE. The solid lines in the lower panels show the average shock radius (left) and explosion energy (right) for the 2D models. For the latter, we show the integrated nuclear energy generation $\dot{\mathcal{E}}_{\mathrm{nuc}}=\int_\mathrm{T<T_\mathrm{th}}{\dot{E}_{\mathrm{nuc}}\,\mathrm{d}V}$ (dotted).}
    \label{fig: Rshock_Eexpl}
\end{figure*}

\begin{figure}[tb]
    \centering
    \includegraphics[width=1.\columnwidth]{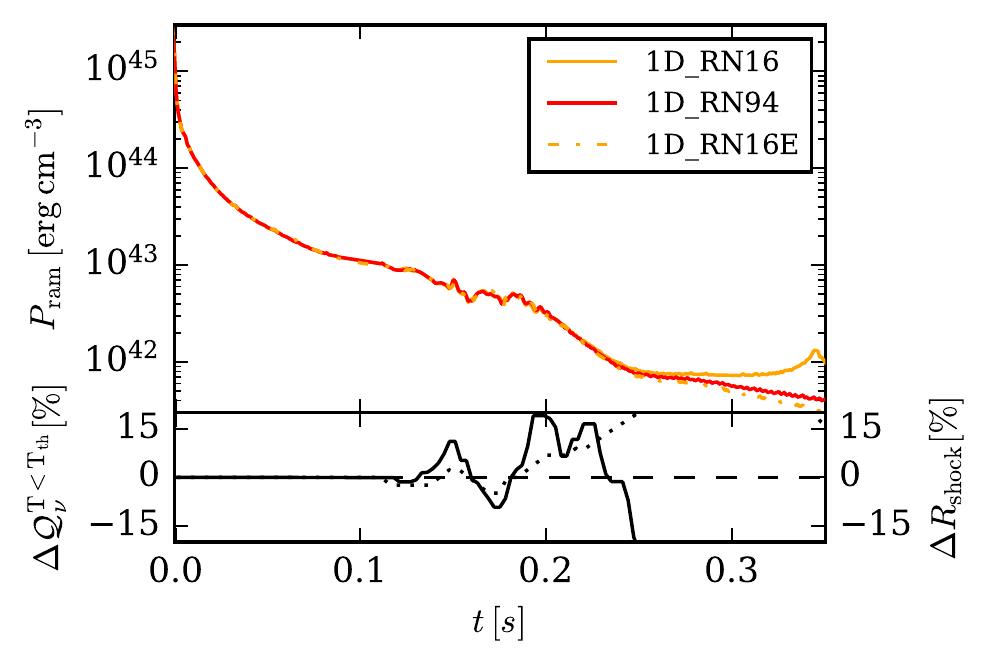}
    \caption{Evolution of the ram pressure for the 1D\_RN94, 1D\_RN16, and 1D\_RN16E models in the first $350 \, \mathrm{ms}$ postbounce. The bottom panel shows the relative differences in the integrated neutrino energy deposition at $T<T_\mathrm{th}$, $\mathcal{Q}^{\mathrm{T<T_\mathrm{th}}}_{\mathrm{\nu}}=\int_\mathrm{T<T_\mathrm{th}}{Q_{\mathrm{\nu}}\,\mathrm{d}V}$, (solid line) and shock radius (dotted) of 1D\_RN94 and 1D\_RN16.}
    \label{fig: RN94_1_vs_rn16_1}
\end{figure}

We start comparing 1D models with different composition treatment, no $Q^\mathrm{LD}_{\mathrm{\nu}}$, and no $\dot{E}_\mathrm{nuc}$ at $T<T_\mathrm{th}$: $\mathrm{1D\_SFHo\_noQ_\nu^{LD}}$, $\mathrm{1D\_flsh\_noQ_\nu^{LD}}$, $\mathrm{1D\_RN16\_noQ_\nu^{LD}}$, and $\mathrm{1D\_RN94\_noQ_\nu^{LD}}$. The upper left panel of Figure~\ref{fig: Rshock_Eexpl} shows the evolution of the shock radius for these models (dotted lines). Without energy transfer between neutrinos and matter at low densities and temperatures, the main variation in the non-SFHo models is the composition, which is an input for the Helmholtz EOS. We note that in the relevant temperature regime the baryonic contribution is negligible, and thus the way in which the nuclear composition is treated has no dynamical effect. Thus, the composition differences between the non-NSE models do not affect the evolution of the shock. However, $\mathrm{1D\_SFHo\_noQ_\nu^{LD}}$ uses a different table for the leptonic contribution than the other three runs. Its slightly different pressure and internal energy modify the evolution of the shock.

When $Q_{\mathrm{\nu}}$ is taken into account in the entire simulation domain (models 1D\_flsh, 1D\_SFHo, 1D\_RN16, and 1D\_RN94, solid lines in the upper left panel of Figure~\ref{fig: Rshock_Eexpl}), the different evolution of the shock is now mainly influenced by this additional energy source term, which dominates the changes between different EOS. 1D\_flsh and 1D\_SFHo have an almost identical behavior due to their very similar composition, which consists of nucleons, alphas (in case of 1D\_SFHo), and a representative nucleus. This leads to a comparable number of nucleons and, thus, similar neutrino opacities and neutrino heating. Note that $Q_\mathrm{\nu}\propto \kappa_\mathrm{a} \propto n_\mathrm{N} \sigma_\mathrm{\nu N} $, where $\kappa_\mathrm{a}$ is the absorption opacity, $n_\mathrm{N}$ is the free nucleon density, and $\sigma_\mathrm{\nu N}$ is the neutrino-nucleon cross section.

Models 1D\_RN16 and 1D\_RN94 show changes in the evolution due to the different species included in the simulation, which changes the number of nucleons, and therefore, the neutrino absorption. The nuclei mapped into simulations employing the RN94 are different than those that are considered in the original progenitor\footnote{ obtained in https://2sn.org/nucleosynthesis/WH2007.shtml}, which provides the aprox19 composition. For this reason, RN94 performs a readjustment where $(\alpha,p)$ reactions are dominant in the oxygen-rich layers. This leads to a difference in the mass fractions of the neutrons of $\Delta \mathrm{n} \equiv \log_{10}{( \frac{X^{\mathrm{RN94}}_\mathrm{n}}{X^{\mathrm{RN16}}_\mathrm{n}})} =0.3$ and $\Delta \mathrm{p} = 2.3$ for protons\footnote{For more differences on nucleons, for a characteristic trajectory, see Figure~\ref{fig:renet_npa} in Appendix~\ref{sec: appendix network}}, which results in an increase in the opacity in the vicinity of the shock. This translates into differences of up to $15 \%$ in the neutrino heating in the low-density regime (Figure~\ref{fig: RN94_1_vs_rn16_1}) and into a decrease in the ram pressure in the shock, which, although small, is sufficient to modify its balance, allowing for an easier expansion in 1D\_RN94.

\subsubsection{Nuclear energy generation in 1D}
\label{subsubsct: engen1_dyn}
\begin{figure}[tb!]
    \centering
    \includegraphics[width=1.\columnwidth]{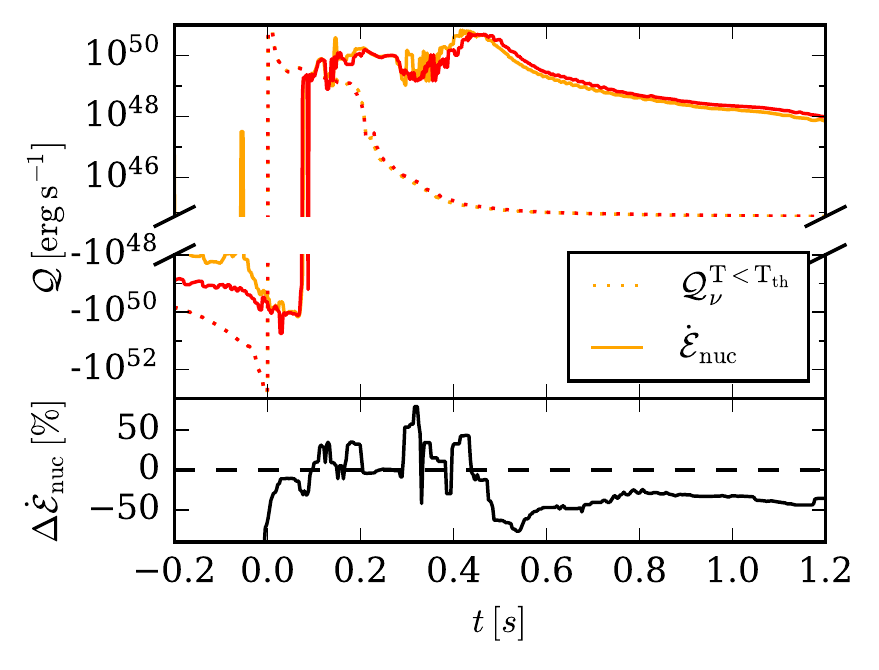}
    \caption{The upper panel shows the evolution of the integrated energy source terms in the $T<T_\mathrm{th}$ region ($\mathcal{Q}$) for the models 1D\_RN94E and 1D\_RN16E in red and yellow, respectively. $\mathcal{Q}^{\mathrm{T<T_\mathrm{th}}}_{\mathrm{\nu}}=\int_\mathrm{T<T_\mathrm{th}}{Q_{\mathrm{\nu}}\,\mathrm{d}V}$ and $\dot{\mathcal{E}}_{\mathrm{nuc}}=\int_\mathrm{T<T_\mathrm{th}}{\dot{E}_{\mathrm{nuc}}\,\mathrm{d}V}$. The lower panel shows the relative difference in the energy released by nuclear reactions in the low-density region.}
    \label{fig:Enuc_Qnu}
\end{figure}

The energy from nuclear reactions significantly changes the dynamics. The middle left panel of Figure~\ref{fig: Rshock_Eexpl} shows the shock evolution for simulations performed with the RN16 and RN94 networks, taking into account the three different configurations for $\dot{E}_\mathrm{nuc}$. The impact of $\dot{E}_\mathrm{nuc}$ on the dynamics depends on whether the energy is released in the progenitor shells or the postshock region. To show it in a clear manner, we first focus on the different evolution in the models with RN16. The absence of $\dot{E}_\mathrm{nuc}$ in the progenitor infalling shells in 1D\_RN16 and 1D\_RN16e leads to a very similar evolution of the shock and the mass shells. The shock is revived after experiencing a fallback at $t\sim300 \, \mathrm{ms}$ and gains enough energy from the system to expand. Nevertheless, in 1D\_RN16e, it expands slightly faster when the nuclear reactions are starting to feed the system with energy. This is better depicted in the middle right panel of Figure~\ref{fig: Rshock_Eexpl}, where the explosion energy shows that model 1D\_RN16e is more than $15\%$ more energetic than 1D\_RN16 after $500 \, \mathrm{ms}$ postbounce due to the nuclear energy released in the explosive nucleosynthesis. 1D models with nuclear energy generation also in the progenitor shells (1D\_RN16E, 1D\_RN94E, and 1D\_flshE) experience an early expansion without fallback, and fewer layers are accreted to the PNS. Initially, the Fe-group-rich layers absorb energy mainly through $(\gamma,\alpha)$, $(n,\alpha)$, and $(\gamma,n)$ reactions, which slightly accelerates the collapse \citep[e.g.,][]{couch17}. The energy released in outer mass shells, i.e. oxygen and silicon layers, is dominated by $(\alpha,\gamma)$ and $(\mathrm{p},\gamma)$ reactions, which produce $\dot{E}_\text{nuc} >0$. This heats up the infalling matter substantially (see Figure \ref{fig:Enuc_Qnu}). The contribution of the nuclear reactions is higher than that of neutrino absorption for $T\leq T_\mathrm{th}$, from $t=0.1 \,\mathrm{s}$ on. Therefore, nuclear reactions supply a significant amount of energy to the infalling layers, more than $10^{50} \, \mathrm{erg \, s^{-1}}$, which is comparable to the change in internal energy ($\sim 10^{50}-10^{51} \, \mathrm{erg \, s^{-1}}$) in the region. This becomes the dominant heating source for $T\leq T_\mathrm{th}$, and further increases the explosion energy. The additional energy source leads to a decrease in the ram pressure on the shock and allows it to expand easier (Figure~\ref{fig: RN94_1_vs_rn16_1}), in agreement with \citet{nakamura2014,bruenn2006}. 

We note that in terms of the shock radius and explosion energy, the evolution of model 1D\_flshE proceeds similarly to that of models 1D\_RN16E and 1D\_RN94E for the first $\sim 300 \, \mathrm{ms}$.  During this period, the nuclear energy generation of the flashing scheme with its approximate composition of $^{28}\mathrm{Si}$ and $^{56}\mathrm{Ni}$ is not too different from that in the network models. At later times, reactions that cannot be represented by these two nuclei become more important and, consequently, the explosion energy of the flashing model starts to deviate from that of the network models (see the middle right panel of Figure~\ref{fig: Rshock_Eexpl}). This behavior seems to indicate that a flashing scheme with more nuclei might reproduce the evolution of the network more closely.

\begin{figure*}[t]
\gridline{\fig{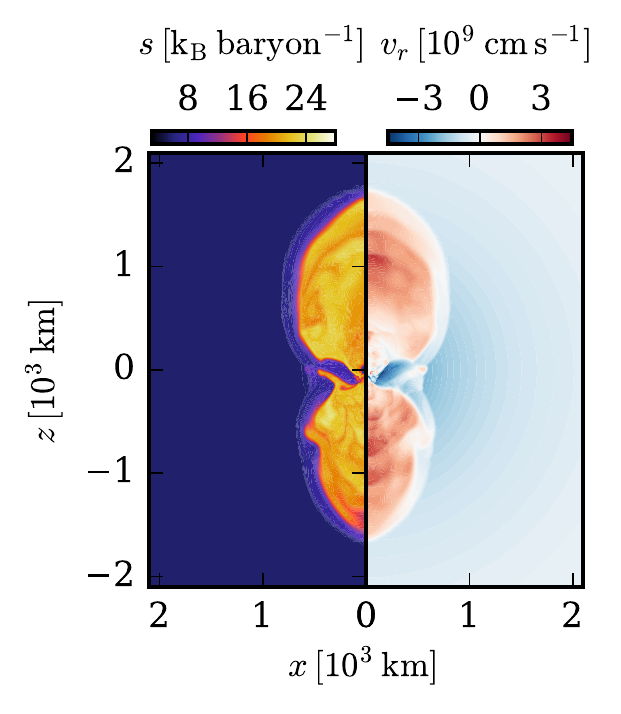}{0.3\textwidth}{(a) 2D\_flsh. $t=500 \, \mathrm{ms}$.\label{fig:sfig1}}
          \fig{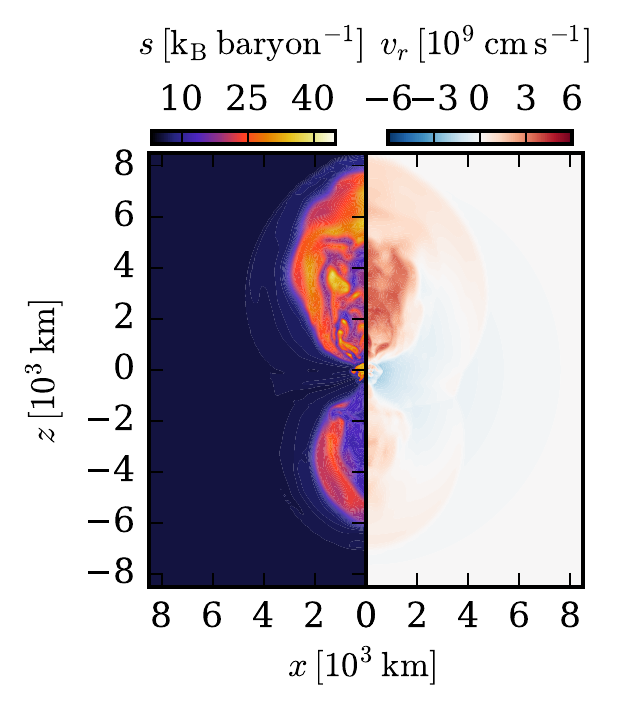}{0.3\textwidth}{(b) 2D\_flsh. $t=1000 \, \mathrm{ms}$.\label{fig:sfig2}}
          \fig{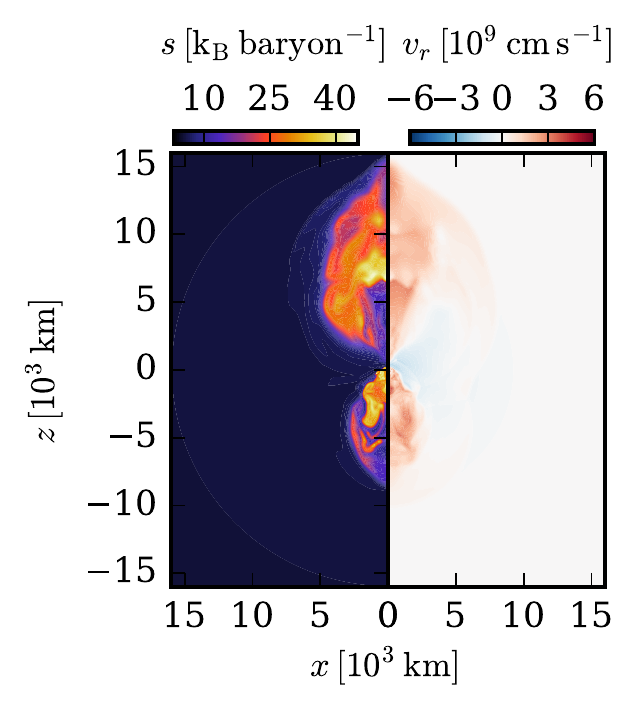}{0.3\textwidth}{(c) 2D\_flsh. $t=1500 \, \mathrm{ms}$.\label{fig:sfig3}}}
\gridline{\fig{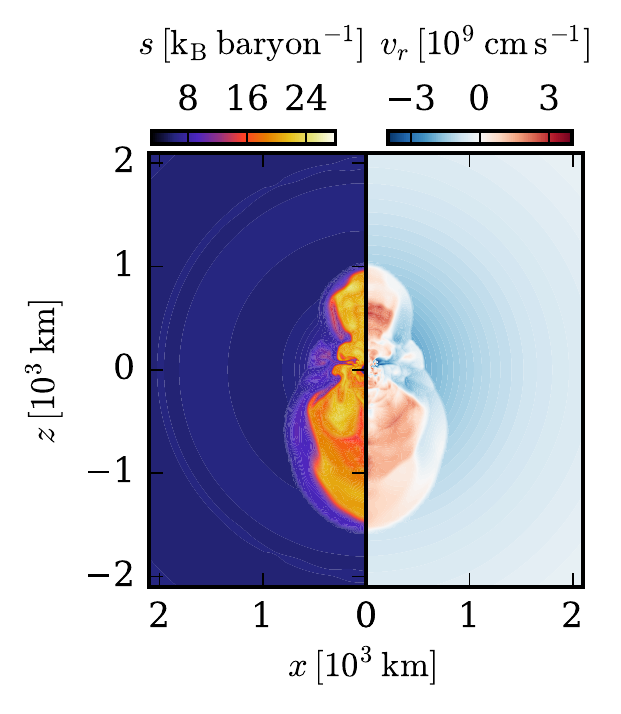}{0.3\textwidth}{(d) 2D\_RN16E. $t=500 \, \mathrm{ms}$.\label{fig:sfig4}}
          \fig{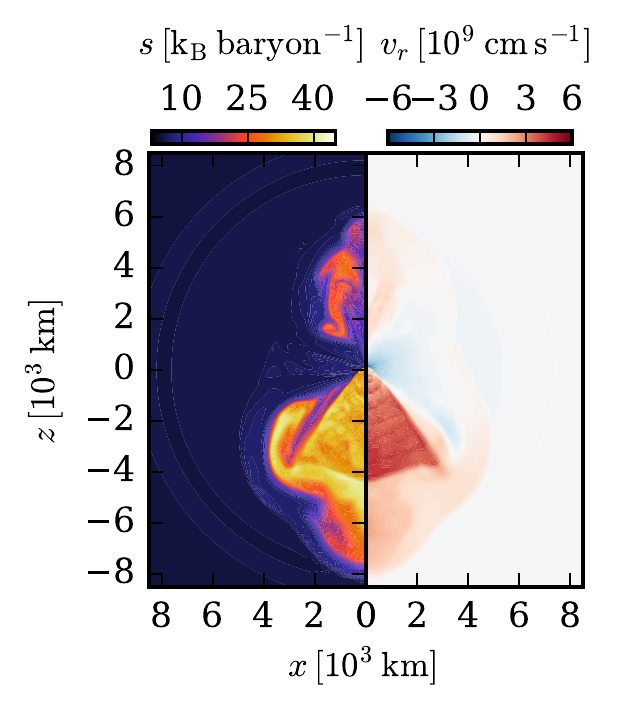}{0.3\textwidth}{(e) 2D\_RN16E. $t=1000 \, \mathrm{ms}$.\label{fig:sfig5}}
          \fig{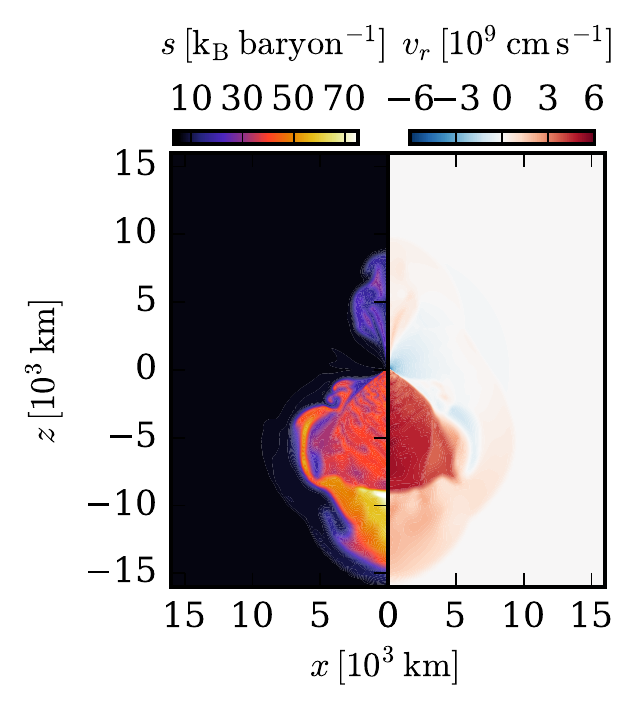}{0.3\textwidth}{(f) 2D\_RN16E. $t=1500 \, \mathrm{ms}$.\label{fig:sfig6}}}
\gridline{\fig{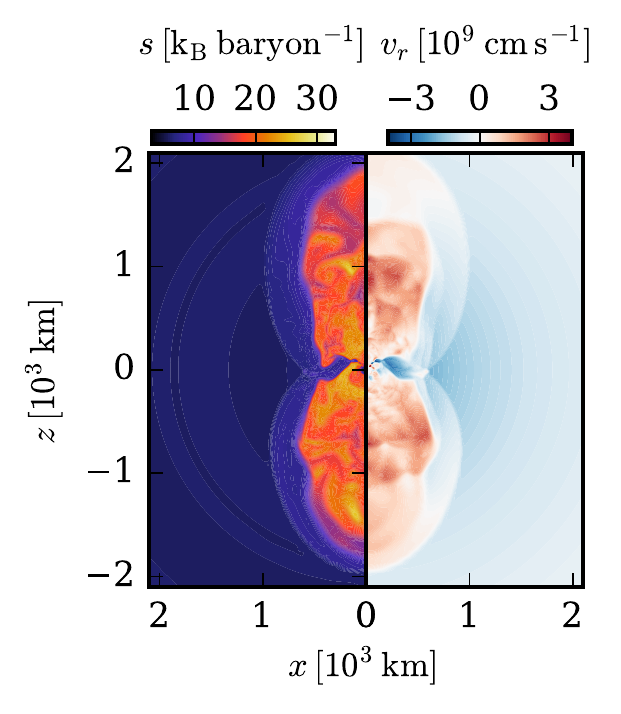}{0.3\textwidth}{(g) 2D\_RN94E. $t=500 \, \mathrm{ms}$.\label{fig:sfig7}}
          \fig{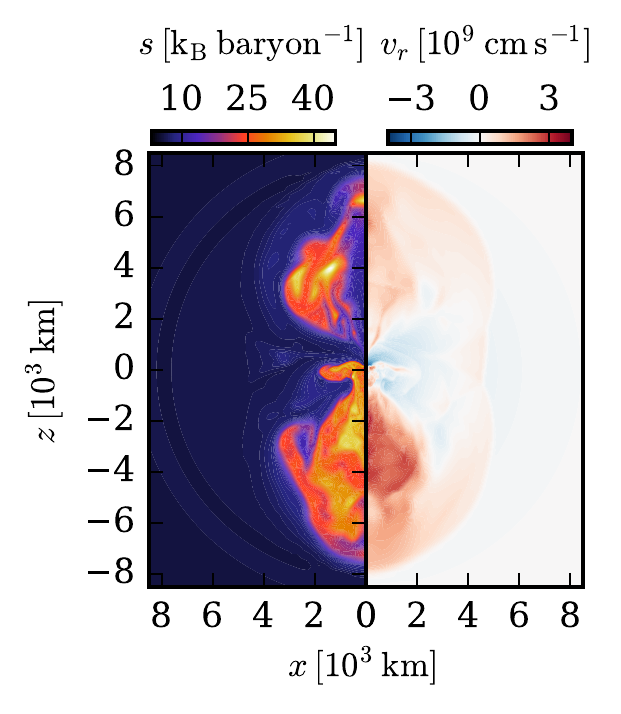}{0.3\textwidth}{(h) 2D\_RN94E. $t=1000 \, \mathrm{ms}$.\label{fig:sfig8}}
          \fig{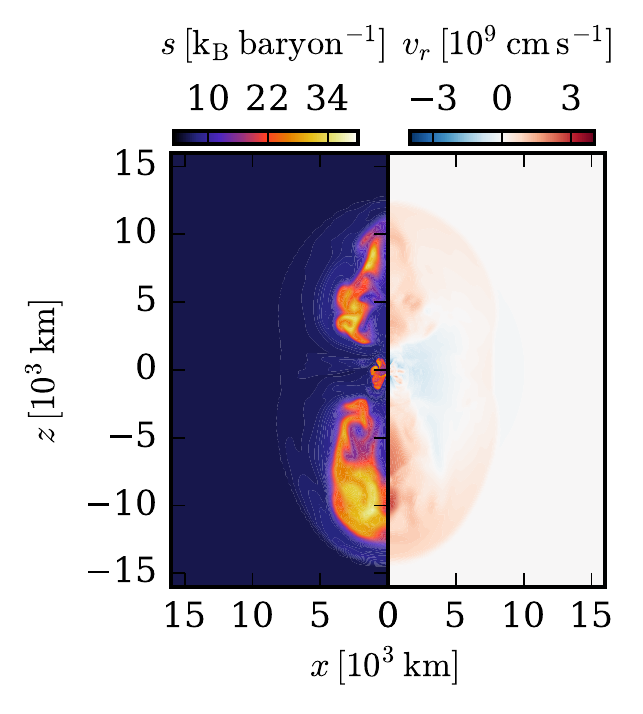}{0.3\textwidth}{(i) 2D\_RN94E. $t=1500 \, \mathrm{ms}$.\label{fig:sfig9}}}
\caption{Slices of entropy and radial velocity in the 2D models at $500 \, \mathrm{ms}$, $1000 \, \mathrm{ms}$, and $1500 \, \mathrm{ms}$ of simulation.}
\label{fig:2d_slides}
\end{figure*}
\begin{figure*}[t!]
\centering
    \includegraphics[width=2.\columnwidth]{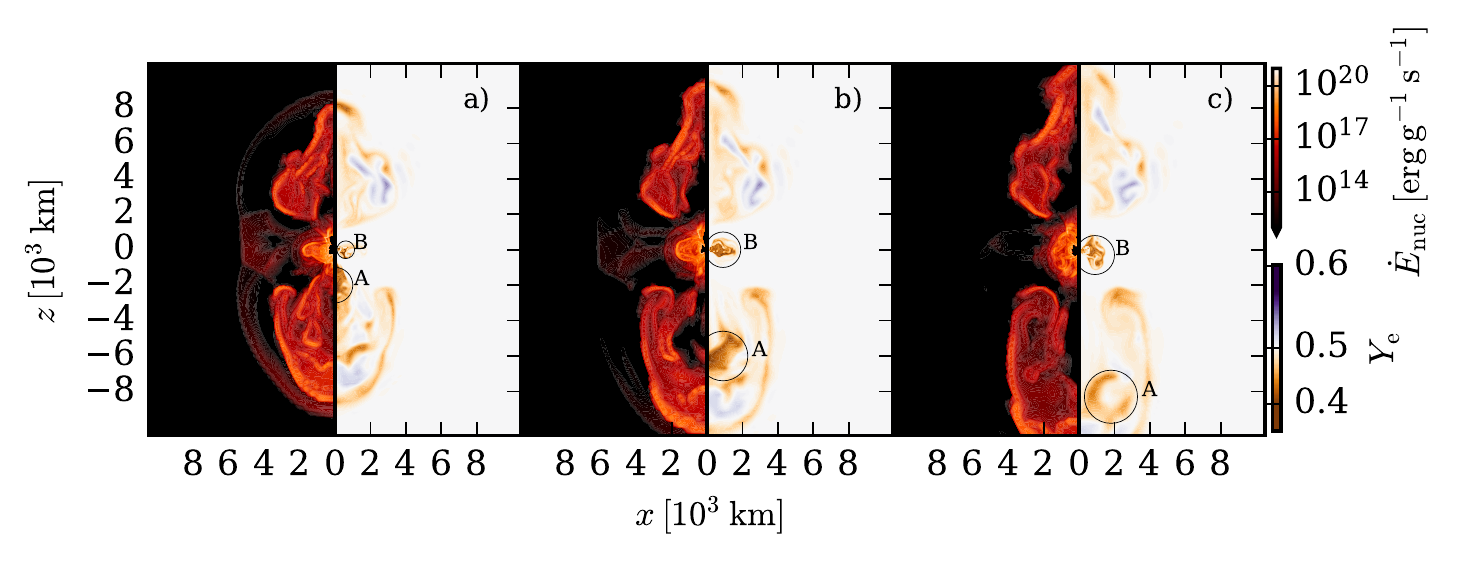}
\caption{Snapshots of the magnitude of $\dot{E}_\text{nuc}$, and $Y_\mathrm{e}$ of model 2D\_RN94E at $t=1100\,\mathrm{ms}$, $t=1250\,\mathrm{ms}$, and $t=1400\,\mathrm{ms}$, respectively. We highlight the different evolution of two low-$Y_\mathrm{e}$ clumps (A and B) ejected from the vicinity of the PNS. While A is ejected in the polar direction and is eventually mixed up with external layers, B cannot overcome the accretion in the equatorial region and stagnates. Nevertheless, the high nuclear energy released in that region helps to sustain this outflow in time.}
\label{fig:engenYe}
\end{figure*}
Despite the similar overall evolution for 1D\_RN16E and 1D\_RN94E, we encounter some differences in the nuclear energy production. Before the bounce, as pointed out previously, the photodisintegration reactions in the iron group absorb energy from the environment. The RN16 is not able to track these nuclei efficiently due to its simplicity. Because of this, we observe a spike of $\dot{\mathcal{E}}_{\mathrm{nuc}}>0$, where $(\mathrm{\alpha},\gamma)$ reactions in the oxygen- and silicon-rich shells are able to release more energy than photodisintegrations absorb in the iron- and nickel-rich layer. In contrast, the RN94 has a broader set of nuclei in the relevant region of the nuclear chart and therefore includes a more complete set of reactions, such as e.g., additional $(n,\gamma)$ reactions. This leads to the observed differences. At intermediate times, between the bounce and $t\sim 500 \, \mathrm{ms}$, $\mathrm{^{24}Mg}(\alpha,\gamma)$, $\mathrm{^{27}Al}(p,\gamma)$, $\mathrm{^{28}Si}(\alpha,\gamma)$, and $\mathrm{^{54}Fe}(p,\gamma)$ are the main contributors to the energy generation in the accreted layers. Because all these reactions are included in RN16 and RN94, the energy production in both networks at these times is comparable, as is shown by the fluctuations of $\Delta \dot{\mathcal{E}}_\mathrm{nuc}$ in Figure~\ref{fig:Enuc_Qnu}. Finally, from $t\sim 500 \, \mathrm{ms}$ on, the nuclear energy is mostly produced in the postshocked region. We observe how $\Delta \dot{\mathcal{E}}_\mathrm{nuc}$ stabilizes around $50\%$. Part of this difference is due to the slightly different shock evolution, which evolves slightly faster in 1D\_RN16E. The remaining discrepancy cannot be easily broken down to an individual reaction, but rather to an overall slightly different nucleosynthetic path. The different number of included nuclei in the calculation causes the differences. While most of the main reactions contributing to $\dot{\mathcal{E}}_{\mathrm{nuc}}$ in this region are included in both networks, e.g,$\mathrm{^{16}O}(\alpha,\gamma)$,  $\mathrm{^{28}Si}(\alpha,\gamma)$, $\mathrm{^{54}Fe}(p,\gamma)$, or $\mathrm{^{55}Co}(p,\gamma)$, additional available paths in RN94 allow for the inclusion of several secondary reactions, which provide additional energy to the system, e.g, $\mathrm{^{58}Ni}(p,\gamma)$ and $\mathrm{^{59}Cu}(p,\gamma)$. Furthermore, a difference in the number of nucleons and $\alpha$ can lead to an additional deviation in the nuclear energy.

\subsubsection{Nuclear energy generation in 2D}
\label{subsubsct: engen2_dyn}
All of the 2D models show typical supernova explosions, with explosion energies (bottom panels of Figure~\ref{fig: Rshock_Eexpl}) of several $10^{50}\, \mathrm{erg}$ and the characteristic prolate shape due to axisymmetry \citep{bruenn16,Summa_2016,Vartanyan_2018}.
The nuclear energy generation also leads to changes in the morphology of the explosion in the 2D models (Figure~\ref{fig:2d_slides}). However, $\dot{E}_\mathrm{nuc}$ is less determinant for the onset of the explosion than in 1D because relaxing spherical symmetry allows for nonradial deformations and convection, which enhance the neutrino heating in the gain layer and trigger an easier shock expansion.

The long-term evolution, in particular the growth of the explosion energy, can differ quite significantly even between otherwise similar models. The shock waves of models 2D\_flsh and 2D\_RN16E expand anisotropically with a moderate north-south asymmetry at early times (left column in Figure~\ref{fig:2d_slides}). The comparable weakness of the shocks in one of the hemispheres allows for important accretion streams toward the PNS. These lead to a significant increase in the neutrino emission (and therefore in $E_\mathrm{exp}$) that trigger an $\mathrm{\alpha}$-rich $\mathrm{\nu}$-driven outflow toward the southern pole (the cone filled with high-entropy high-velocity gas in Figure~\ref{fig:2d_slides}e) in 2D\_RN16E and an axial outflow in 2D\_flsh. 
In 2D\_RN16E, the effect of $\dot{E}_\mathrm{nuc}$ on $R_\mathrm{shock}$ is small and contributes to accelerating it around $t\sim 800 \, \mathrm{ms}$ after bounce. The rate at which nuclear reactions deposit energy in the wind and postshock region is significant, comparable to the growth rate of $E_\mathrm{exp}$ (lower right panel of Figure~\ref{fig: Rshock_Eexpl}). This supports the outflows and leads to a more energetic explosion.

The shock in 2D\_RN94E, in contrast, initially expands more symmetrically toward higher radii (bottom right panel of Figure~\ref{fig: Rshock_Eexpl}), which we attribute to the higher amount of energy generated by nuclear reactions around $t = 100 \, \mathrm{ms}$. This reduces the accretion inflows toward the PNS as a comparison of the masses indicates: $M_\mathrm{PNS}=1.67,\, 1.72$, and $1.71\, \mathrm{M_\odot}$ for 2D\_RN94E, 2D\_flsh, and 2D\_RN16E, respectively.
Therefore, the number of neutrinos emitted after $t\sim 500\,\mathrm{ms}$ is significantly lower, which explains the absence of $\nu\mathrm{-driven}$ winds in 2D\_RN94E and the stagnation of $E_\mathrm{exp}$, in contrast to 2D\_RN16E and 2D\_flsh.

At later times, downflows accreting onto the PNS squeeze its polar region, launching high velocity neutron-rich outflows ($Y_\mathrm{e} \sim 0.35$) ejected from the vicinity of the PNS. This is shown for model 2D\_RN94E in Figure~\ref{fig:engenYe}, where we observe a neutron-rich bubble (indicated by the circle A) that moves from the center (panel~a) to $z\sim-6000\,\mathrm{km}$ in only $150\,\mathrm{ms}$ (panel~b), and continues evolving toward higher latitudes, mixing with the surrounding material (panel~c). The high $v_\mathrm{r}$, up to $\sim 0.16 \, c$, avoid excessive neutrino absorption that would increase the electron fraction for the conditions prevailing here. These outflows tend to be amplified by the nuclear energy generation in the already shocked areas. The predominantly polar direction of this bubble may at least partly be due to the assumed axisymmetry rather than being a general feature of such neutron-rich clumps. In fact, the presence of another such bubble (circle B in Figure~\ref{fig:engenYe}) shows that they can be formed at any latitude as a consequence of the random dynamics of convection and accretion streams in the vicinity of the PNS. The lower radial velocity of the equatorial outflow ($\sim 0.07 \, c$) and the accretion of the upper layer prevent it from evolving toward a larger radius. Heating by nuclear energy reactions at rates around $\dot{E}_\mathrm{nuc}\sim 10^{20} \, \mathrm{erg}\;\mathrm{g}^{-1}\mathrm{s}^{-1}$ plays an important role in sustaining the clump. As we show in the next section, this mechanism allows for the production of more neutron-rich species.

To sum up, in Section~\ref{sct:imp_dynamics} we have studied the impact of reduced networks on the dynamics of the explosion in both 1D and 2D models. The different nuclei included can modify the number of nucleons and hence the neutrino absorption at $T < T_{\mathrm{th}}$. This different heating in the infalling progenitor shells has an impact on the ram pressure and can therefore change the shock evolution. Analogously, the nuclear energy released in that region allows for a decrease of the ram pressure on the shock and favors an easier expansion. In contrast, in the shocked region, nuclear reactions are able to increase the explosion energy substantially. We demonstrated this by comparing the explosion energy of the models 1D\_RN16 and 1D\_RN94 ($\dot{E}_\mathrm{nuc}=0$) to the explosion energy of the models 1D\_RN16e and 1D\_RN94e ($\dot{E}_\mathrm{nuc}$ only in the shocked region). The increase in $E_\mathrm{exp}$ when $\dot{E}_\mathrm{nuc}$ starts to release in the shocked region is striking for both cases (see the solid vs dashed lines in Figure~\ref{fig: Rshock_Eexpl}). The 2D\_RN94 model suggests that the important nuclear energy generation in this region helps to sustain late low-$Y_\mathrm{e}$ outflows in the equatorial direction. Finally, RN16 and RN94 show small differences in $\dot{E}_\mathrm{nuc}$.

\begin{figure*}[t!]
    \centering
    \includegraphics[width=2.\columnwidth]{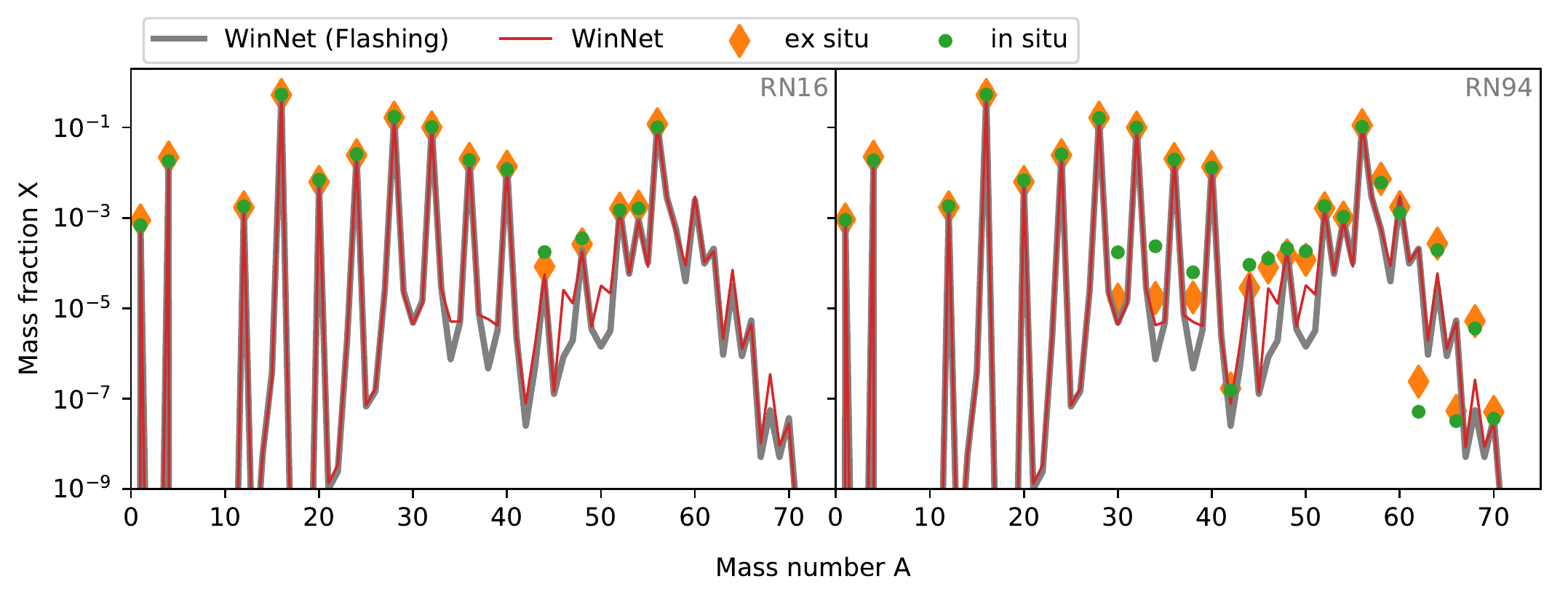}
    \caption{ Integrated final ejecta composition of the 1D models. The red lines correspond to the composition obtained in postprocessing with the full network \textsc{WinNet} in the 1D\_RN16E (left) and 1D\_RN94E (right). The postprocessing results for the 1D\_flsh are displayed in gray for comparison. The green dots stand for the values obtained from the network in situ, i.e. evolved in the simulation. The values obtained with the same reduced network in postprocessing are depicted by orange diamonds.}
    \label{fig:Xi_integrated_1d}
\end{figure*}
\begin{figure*}[t!]
    \centering
    \includegraphics[width=2.\columnwidth]{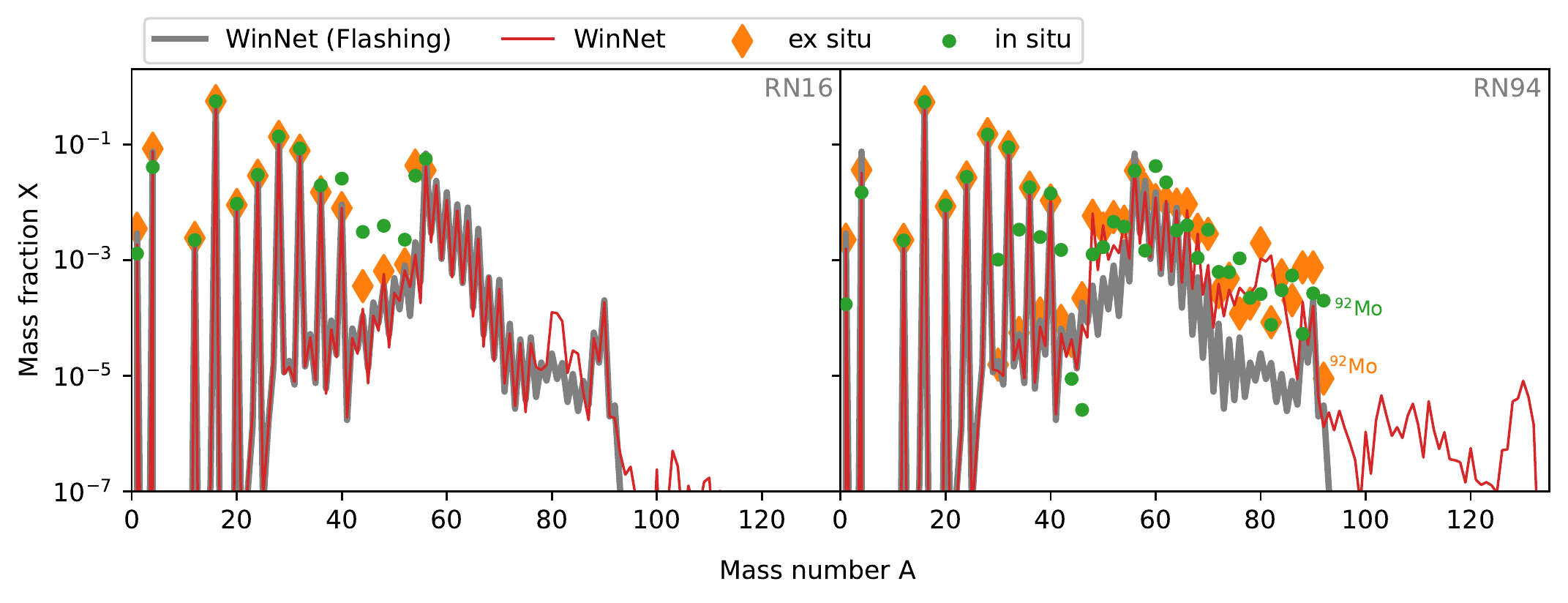}
    \caption{Integrated final composition of the 2D models. Analogously to Figure~\ref{fig:Xi_integrated_1d}, the gray line corresponds to the postprocessing calculation of model 2D\_flsh.}
    \label{fig:Xi_integrated_2d}
\end{figure*}
\subsection{Impact on the nucleosynthesis}

In this section, we study the effects that reduced networks have on the composition of the ejecta at the end of the simulation, i.e., $t=1.5\,\mathrm{s}$, in 1D and 2D models.
\subsubsection{Postprocessing with \textsc{WinNet}}

First, we show the impact on the abundances by computing them with the postprocessing nucleosynthesis network \textsc{WinNet}. We compare the abundances obtained in models that include nuclear energy generation (1D\_RN16E, 1D\_RN94E, 2D\_RN16E, and 2D\_RN94E) with those from models that do not take $\dot{E}_\text{nuc}$ into account (1D\_flsh and 2D\_flsh) in 1D and 2D, respectively.
 
In spherically symmetric models, $Y_\mathrm{e}$ is very close to $0.5$. Therefore, species beyond the iron group are not synthesized (Figure~\ref{fig:Xi_integrated_1d}). This also explains the only minor differences in the postprocessing final composition with \textsc{WinNet} for the 1D\_flsh, the 1D\_RN16E, and 1D\_RN94E models, despite the different dynamic evolution when the energy generation from the nuclear reactions is taken into account.

In the 2D models (Figure~\ref{fig:Xi_integrated_2d}), a larger number of nuclei are involved due to the broader range of $Y_\mathrm{e}$. The 2D\_flsh and 2D\_RN16E model show standard CCSN nucleosynthesis, where mainly iron-group elements are formed together with lighter heavy species around $A\sim90$ \citep[see, e.g.,][]{Eichler_2017,harris17,wanajo18,witt_paper21}. In contrast, model 2D\_RN94E produces larger amounts of heavier species. The composition obtained with \textsc{WinNet} shows a significant enhancement around the first r-process peak, $A\sim80$, particularly $\mathrm{^{84}Ge}$. Moreover, a small amount is observed up to the second r-process peak, $A\sim 130$. This larger production of heavier species takes place in the late low-$Y\mathrm{_e}$ outflows, which are supported by the nuclear energy release as outlined in Section~\ref{sct:imp_dynamics}. We observe an agreement between the postprocessing RN94 (ex situ) and \textsc{WinNet} abundances in the 2D\_RN94E model until $A\sim90$, showing that RN94 is able to reproduce the main nuclei that are synthesized in CCSN. There is a difference in $\mathrm{^{92}Mo}$ that can be explained by the fact that it acts as a bottleneck for heavier species in RN94.  
\vspace{10pt}

\begin{figure*}[t!]
    \centering
    \includegraphics[scale=1.]{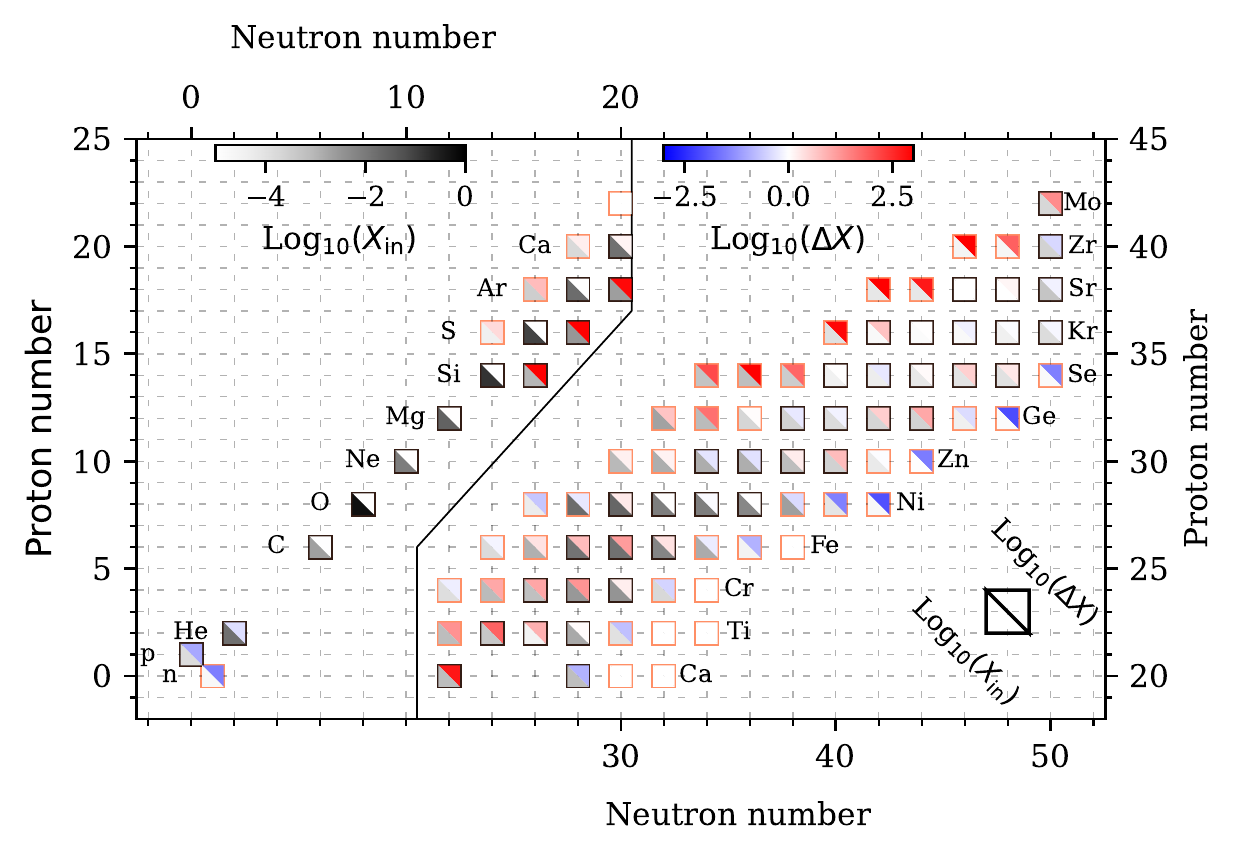}
    \caption{Chart with the RN94 isotopes in boxes. The orange edges indicate unstable nuclei, and black edges show stable nuclei. The bottom half of the boxes depict in situ integrated mass fractions for 2D\_RN94E at the end of the simulation. The upper half show the differences with respect to the ex situ mass fraction, defined as $\Delta X= \frac{X_\mathrm{in}}{X_\mathrm{ex}}$, for species with $X_\mathrm{i}>10^{-5}$.}
    \label{fig:insitu_vs_exsitu}
\end{figure*}

\subsubsection{Ex situ vs in situ calculations}
In this section, we show the differences of computing the nucleosynthesis in the simulation (in situ) and in postprocessing (ex situ). For the comparison, we calculate the latter with the same network employed in the hydro (RN16 or RN94, depending on the model).

In Figures~\ref{fig:Xi_integrated_1d}~and~\ref{fig:Xi_integrated_2d}, we also show the composition obtained in situ in the models with the networks and its respective postprocessing, ex situ calculation. The goal is to analyze which differences come from evolving the network together with the hydro. The 1D\_RN16E and 1D\_RN94E models show very good agreement in the most abundant species, such as $\mathrm{^{56}Ni}$. However, there are some discrepancies for $\mathrm{^{30}S}$, $\mathrm{^{34}Ar}$, $\mathrm{^{40}Ca}$, $\mathrm{^{44}Ti}$, $\mathrm{^{52}Cr}$, and $\mathrm{^{62}Zn}$ with $\log_{10}(\Delta X) \equiv \log_{10}{( \frac{X_\mathrm{in}}{X_\mathrm{ex}})} \approx 0.5-1$. Lagrangian and Eulerian methods can both suffer from numerical errors, although in different ways. Lagrangian tracer particles, used to compute the composition in postprocessing, are more uncertain when tracking low-density ejecta and especially the products of the $\mathrm{\alpha}$-rich freeze-out, as shown in \citet{harris17}. Furthermore, accurately representing features varying on short length and time scales such as those arising from multidimensional potentially turbulent flows can be difficult and require an exceedingly large number of tracers. The numerical diffusivity that Eulerian methods require for stability can lead to artificial mixing of species across the grid. While its importance decreases with increasing grid resolution, much finer grids than feasible are necessary to suppress this effect. This is the origin of the systematic disagreement between in situ and ex situ calculations in our models. 

In 1D, the representation of the trajectories is much simpler and accurate, reducing the uncertainty to that arising mainly from the aforementioned numerical diffusion. We therefore tested its impact based on the 1D\_RN94E model by recalculating it with different resolutions of $500$, $600$, $700$, and $1000$ radial zones. While the ex situ abundances for all models were almost identical, with maximum deviations of the order of $\sim 40\%$ for $^{42}\mathrm{Ti}$ and $^{54}\mathrm{Ni}$, some elements (i.e.,  $\mathrm{^{30}S}$, $\mathrm{^{34}Ar}$, and $\mathrm{^{62}Zn}$) showed variations in the order of $\sim 60\%-75\%$ for the in-situ abundances. All other elements also converged within the in-situ networks. The impact of numerical diffusion could be decreased in further studies by modifying the advection scheme of the composition \citep{plewa_difussion}. 

The 2D\_RN16E model has differences on the same order as the previous ones in $\mathrm{^{40}Ca}$, $\mathrm{^{44}Ti}$, $\mathrm{^{48}Cr}$, and $\mathrm{^{52}Fe}$. The overproduction of these isotopes in situ is consistent with the explosion morphology seen in Figure~\ref{fig:2d_slides}. It is characterized by an extended high-entropy and high-velocity shocked region in which mainly $\mathrm{\alpha}$-rich freeze-out takes place. The in situ and ex situ discrepancies in 2D\_RN94E are depicted in Figure~\ref{fig:insitu_vs_exsitu}. Some differences in the overall nucleosynthesis trend can be observed by looking at the most neutron-rich and deficient isotopes of Zn, Ge, Se, Kr, Sr, and Zr. The postprocessing underproduction of $\mathrm{^{44}Ti}$, $\mathrm{^{46}Ti}$, $\mathrm{^{48}Ti}$, $\mathrm{^{48}Cr}$, $\mathrm{^{50}Cr}$, $\mathrm{^{52}Cr}$, $\mathrm{^{54}Fe}$, and $\mathrm{^{56}Fe}$ predominantly leads to a more neutron-rich path with bottlenecks in $\mathrm{^{80}Ge}$ and $\mathrm{^{84}Se}$. In situ abundances show a tendency of being more neutron-deficient, starting from the aforementioned isotopes of $\mathrm{Ti}$, $\mathrm{Cr}$, and $\mathrm{Fe}$. Figure~\ref{fig:insitu_vs_exsitu} shows that matter accumulates at the proton-rich isotopes $\mathrm{^{72}Se}$, $\mathrm{^{74}Se}$, and $\mathrm{^{76}Se}$, suggesting that even the highly extended RN94 network still has some bottlenecks that could be better represented with an even more extended network. Finally, this more neutron-deficient path ends up producing more $\mathrm{^{92}Mo}$, $\log_{10}(\Delta X_\mathrm{^{92}Mo}) = 1.35$.

\section{Summary \& Conclusions}
\label{sec: conclusions}

We have presented a detailed study of how the treatment of the composition within CCSN simulations impacts the explosion dynamics and nucleosynthesis.

We performed 1D and 2D CCSN simulations using the neutrino-hydrodynamics code \textsc{Aenus-Alcar} \citep{Aenus-Alcar-Just,Just__2018__MonthlyNoticesoftheRoyalAstronomicalSociety__CoreCollapseSupernovaSimulationsinOneandTwoDimensionsComparisonofCodesandApproximations,martin_aloy_magnetorotational}.  So far, this code included the nuclear reactions outside the NSE regime only via the simplified flashing scheme \citep{flashing-scheme}, which assumes that the gas consists only of nucleons and a representative nucleus, for which, depending on the temperature, we use $\mathrm{^{28}Si}$ or $\mathrm{^{56}Ni}$. We used the reduced network module \textsc{ReNet} (see Appendix~\ref{sec: appendix network}) to replace the flashing scheme by a $16\,\mathrm{\alpha}$-chain (RN16) and a $94$-isotope network (RN94).  The latter is able to reproduce the main nucleosynthesis yields in standard CCSN explosions \citep[e.g,][]{Eichler_2017}. In addition, thanks to the $148$ nuclei considered in steady-state approximation, RN94 is the most extended network in the nuclear chart ever employed \footnote{To date of submission. It should also be noted that in terms of the number of isotopes, the one of the Oak Ridge group \citep[e.g.,][]{bruenn20} is larger, including 160 species.} in state-of-the-art hydrodynamic simulations. Both in-situ networks return the composition of the gas and the rate at which nuclear reactions generate or consume internal energy.

The different compositions in the low-density region have an impact on the number of nucleons, which can change the neutrino heating in the vicinity of the shock. This modifies the ram pressure outside of it, and therefore, its evolution. 

We have demonstrated how the energy released in the nuclear reactions impacts the dynamics of the explosion. The energy generation in the preshocked collapsing matter again decreases the ram pressure outside the shock and allows it to expand easier, in agreement with \citet{bruenn2006} \& \citet{nakamura2014}. The nuclear energy released in the shocked region has a significant contribution, up to $20\%$, to the total explosion energy. The flashing scheme with $^{28}\mathrm{Si}$ and $^{56}\mathrm{Ni}$ is not able to reproduce the nuclear energy generation in this region, which leads 1D\_flshE to a lower explosion energy than 1D\_RN16E and 1D\_RN94E. Nevertheless, their overall evolution is similar. The differences between RN16 and RN94 are small regarding the nuclear energy generation, where $(\alpha,\gamma)$ and $(\mathrm{p},\gamma)$ are the main production channels. While the models we presented are not very energetic, we explored more energetic explosions, $E_\mathrm{exp}\sim 1 \, \mathrm{B}$, and obtained a similar impact.

Finally, we obtained the detailed nuclear yields of the models by applying the nuclear network \textsc{WinNet} with $6545$ isotopes in an exsitu postprocessing step to Lagrangian tracer particles tracking the fluid flow.  We compared the postprocessing results among different models and to the in-situ networks. In 1D, the differences are small because the $Y_\mathrm{e}$ involved are very similar among the models and are close to $0.5$. In 2D, the variation in abundances among different models becomes stronger (Figure~\ref{fig:Xi_integrated_2d}). The energy released in the nuclear reactions helps to sustain late neutron-rich outflows ejected from the vicinity of the PNS. Model 2D\_RN94E shows how this mechanism allows weak r-process to take place. Moreover, we have compared the final composition obtained in situ and ex situ making use of RN16 and RN94. In agreement with \citet{harris17}, we find significant discrepancies mainly in products of the $\mathrm{\alpha}$-rich freeze-out because Lagrangian tracer particles involve larger uncertainties when tracking such regions. For the in situ results, we identify the resolution-dependent numerical diffusion of species with low abundances as a factor contributing to the discrepancies. Moreover, we demonstrate how these uncertainties propagate, leading to variations in the nucleosynthesis path that alter the final yields. 

What are the advantages and disadvantages of evolving a network in the simulations? While Lagrangian tracer methods and the ex situ results based on them are well suited for dense regions \citep[e.g.,][]{price_federrath} and avoid the excess numerical diffusion that may beset grid-based Eulerian schemes and, consequently, in situ abundances, they lack mixing and are more uncertain in tracking low-density regions and their nucleosynthesis, e.g. products of the $\alpha \mathrm{-rich}$ freeze-out. This work suggests that it is necessary to employ in situ realistic networks in CCSN simulations with a fine grid resolution, or with a less numerically diffusive advection scheme, to obtain a realistic feedback of the energy generation, the neutrino opacities, and a more accurate ejecta composition. Thus, this study showed the strengths and weaknesses of employing networks in CCSN simulations and, hopefully, can help future simulations to decide depending on the goal of the study.

\vspace{10pt}
We thank J. Austin Harris, W. Raphael Hix, Finia Jost, and Andre Sieverding for useful discussions. This work was supported by the Deutsche Forschungsgemeinschaft (DFG, German Research Foundation) - Project-ID 279384907 - SFB 1245, the European Research Council under grant EUROPIUM-667912, and the State of Hessen within the Research Cluster ELEMENTS (Project ID 500/10.006). The authors gratefully acknowledge the computing time provided to them on the high-performance computer Lichtenberg at the NHR Centers NHR4CES at TU Darmstadt (Project-ID 1527). This is funded by the Federal Ministry of Education and Research, and the state governments participating. MR and MO acknowledge support through the grant PID2021-127495NB-I00 funded by MCIN/AEI/10.13039/501100011033 and by the European Union, and the Astrophysics and High Energy Physics programme of the Generalitat Valenciana ASFAE/2022/026 funded by MCIN and the European Union NextGenerationEU (PRTR-C17.I1). MR acknowledges support from the Spanish Ministry of Science via the Juan de la Cierva programme (FJC2021-046688-I). MO acknowledges support from the Spanish Ministry of Science via the Ram\'on y Cajal programme (RYC2018-024938-I).

\bibliography{final.bib}

\appendix
\section{ReNet: Reduced Network}
\label{sec: appendix network}
We developed a highly flexible and adaptable reduced nuclear reaction network code, \textsc{ReNet}. This reaction network solves the nuclear reaction network equations \citep[e.g.,][]{clayton68,winteler12a,Hix-thielemann1999,Hix2006,lippuner17a,Cowan2021}:
\begin{align}\label{eq:nuclear_reaction_network}
\dot{Y}_i &= \sum _j \mathcal{N}_j ^i \lambda _j Y_j && \text{(Decays, Photodisintegration)} \nonumber \\
&+\sum _{j,k} \frac{\mathcal{N}^i _{j,k}}{1+\delta _{jk}}\rho N_{\mathrm{A}} \langle \sigma v \rangle _{j,k}Y_jY_k  && \text{(2-body)} \nonumber \\
&+\sum _{j,k,l}\frac{\mathcal{N}^i_{j,k,l}}{1+\Delta_{jkl}}\rho ^2N_{\mathrm{A}}^2\langle \sigma v\rangle _{j,k,l} Y_j Y_k Y_l. && \text{(3-body)},
\end{align}
with the abundance $Y_i$ of a nucleus $i$, the number of destroyed or created nuclei within the reaction $\mathcal{N}$, the density $\rho$, the decay constant $\lambda$, and the cross section $\langle \sigma v \rangle$. The factors $\delta_{jk}$ and $\Delta_{jkl} = \delta _{jk}+\delta _{kl}+\delta _{jl}+2 \delta _{jkl}$ ensure that there is no double counting of identical nuclei involved in the equation. This set of coupled differential equations tends to be stiff due to the large range of magnitudes of the reaction rates \citep[e.g.,][]{Hix-thielemann1999}. The system is therefore solved with an implicit Euler algorithm \citep[e.g.,][]{lippuner17a} to ensure numerical stability.


Every considered nucleus in Eq.~\eqref{eq:nuclear_reaction_network} introduces an additional differential equation. However, not all nuclei are connected with all others, and the resulting Jacobian is therefore sparse \citep[e.g.,][]{Hix-thielemann1999,lippuner17a}. We tested whether the computational overhead of a sparse solver (\textsc{PARDISO}, \citealt{INTELPARDISO}) is lower than that of a dense matrix solver (\textsc{LAPACK}, \citealt{lapack}). We find a scaling of $\sim O(N^{1.6})$ and $\sim O(N^{2.8})$ for a sparse and dense solver, respectively. The overhead of the sparse solver makes this approach faster when we consider more than $\sim 300$ nuclei (see, to $N\gtrsim 100$ in \citealt{winteler12a}). For the small networks considered, we therefore use the dense \textsc{LAPACK} solver. 

To reduce the set of considered nuclei without neglecting important reactions, we assume a steady state for some chosen nuclei,
\begin{equation}
    \frac{\mathrm{d}Y_i}{\mathrm{d}t} = 0.
\end{equation}
The strategy of using a steady-state flow through intermediate nuclei was already applied in other reduced networks\footnote{accessed via \url{https://cococubed.com/code_pages/burn_helium.shtml}} such as, e.g., the aprox13 \citep[][]{Timmes1999}, aprox19 \citep[][]{approx19}, and aprox21 \citep[][]{Paxton2011}. As an example of this approximation, consider the three nuclei $^{24}$Mg, $^{28}$Si, and $^{27}$Al. $^{24}$Mg and $^{28}$Si are connected by an $\alpha$-capture. Another possibility to synthesize $^{28}$Si from $^{24}$Mg is given by the reaction chain $^{24}\mathrm{Mg}(\alpha,\mathrm{p})^{27}\mathrm{Al}(\mathrm{p},\gamma)^{28}\mathrm{Si}$. The set of differential equations (not showing protons and alphas) is given by
\begin{align}
    \frac{\mathrm{d}Y_\mathrm{^{24}Mg}}{\mathrm{d}t} = &+\langle \sigma v \rangle _{^{28}\mathrm{Si}(\gamma,\alpha)^{24}\mathrm{Mg}}\, Y_{^{28}\mathrm{Si}} \label{eq:mg_eq}\\ \nonumber
    &- \rho N_\mathrm{A} \langle \sigma v \rangle _{^{24}\mathrm{Mg}(\alpha,\gamma)^{28}\mathrm{Si}}\,Y_{^{24}\mathrm{Mg}}\,Y_{\alpha} \\ \nonumber
       &- \rho N_\mathrm{A} \langle \sigma v \rangle _{^{24}\mathrm{Mg}(\alpha,\mathrm{p})^{27}\mathrm{Al}}\,Y_{^{24}\mathrm{Mg}}\,Y_{\alpha} \nonumber \\ 
       &+ \rho N_\mathrm{A} \langle \sigma v \rangle _{^{27}\mathrm{Al}(\mathrm{p},\alpha)^{24}\mathrm{Mg}}\,Y_{^{27}\mathrm{Al}}\,Y_\mathrm{p} \nonumber \\ 
           \frac{\mathrm{d}Y_\mathrm{^{27}Al}}{\mathrm{d}t} = &+\langle \sigma v \rangle _{^{28}\mathrm{Si}(\gamma,\mathrm{p})^{27}\mathrm{Al}}\, Y_{^{28}\mathrm{Si}} \\ 
    &+ \rho N_\mathrm{A} \langle \sigma v \rangle _{^{24}\mathrm{Mg}(\alpha,\mathrm{p})^{27}\mathrm{Al}}\,Y_{^{24}\mathrm{Mg}}\,Y_{\alpha} \nonumber \\ 
       &- \rho N_\mathrm{A} \langle \sigma v \rangle _{^{27}\mathrm{Al}(\mathrm{p},\alpha)^{24}\mathrm{Mg}}\,Y_{^{27}\mathrm{Al}}\,Y_\mathrm{p} \nonumber \\ 
       &- \rho N_\mathrm{A} \langle \sigma v \rangle _{^{27}\mathrm{Al}(\mathrm{p},\gamma)^{28}\mathrm{Si}}\,Y_{^{27}\mathrm{Al}}\,Y_\mathrm{p} \nonumber \\ 
           \frac{\mathrm{d}Y_\mathrm{^{28}Si}}{\mathrm{d}t} = &-\langle \sigma v \rangle _{^{28}\mathrm{Si}(\gamma,\mathrm{p})^{27}\mathrm{Al}}\, Y_{^{28}\mathrm{Si}} \label{eq:si_eq}\\ 
    &-\langle \sigma v \rangle _{^{28}\mathrm{Si}(\gamma,\alpha)^{24}\mathrm{Mg}}\, Y_{^{28}\mathrm{Si}} \nonumber \\ 
    &+ \rho N_\mathrm{A} \langle \sigma v \rangle _{^{24}\mathrm{Mg}(\alpha,\gamma)^{28}\mathrm{Si}}\,Y_{^{24}\mathrm{Mg}}\,Y_{\alpha} \nonumber \\ 
    &+ \rho N_\mathrm{A} \langle \sigma v \rangle _{^{27}\mathrm{Al}(\mathrm{p},\gamma)^{28}\mathrm{Si}}\,Y_{^{27}\mathrm{Al}}\,Y_\mathrm{p}. \nonumber
\end{align}
Setting $\frac{\mathrm{d}Y_\mathrm{^{27}Al}}{\mathrm{d}t} = 0$, we obtain
\begin{align}
    Y_{^{27}\mathrm{Al}} = &\frac{\langle \sigma v \rangle _{^{28}\mathrm{Si}(\gamma,\mathrm{p})^{27}\mathrm{Al}}\, Y_{^{28}\mathrm{Si}}+\rho N_\mathrm{A} \langle \sigma v \rangle _{^{24}\mathrm{Mg}(\alpha,\mathrm{p})^{27}\mathrm{Al}}\,Y_{^{24}\mathrm{Mg}}\,Y_{\alpha}}%
    {\rho \, N_\mathrm{A} \left(\langle \sigma v \rangle _{^{27}\mathrm{Al}(\mathrm{p},\alpha)^{24}\mathrm{Mg}} +\langle \sigma v \rangle _{^{27}\mathrm{Al}(\mathrm{p},\gamma)^{28}\mathrm{Si}} \right)Y_\mathrm{p}},
\end{align}
which can be inserted into Eq.~\eqref{eq:mg_eq} and \eqref{eq:si_eq}, eliminating the necessary time evolution of $Y_\mathrm{^{27}Al}$. We applied this procedure to a series of nuclei using the \textsc{Sympy} Python package to derive the equations and the Jacobian analytically. The latter is necessary because proton or neutron abundances can end up in the denominator when we include all reactions that go through the steady-state nucleus. In this way, we are able to evolve a large region of the nuclear chart without explicitly evolving the abundance of all nuclei. 

\begin{figure}[tb!]
    \centering
 \includegraphics[width=0.5\columnwidth]{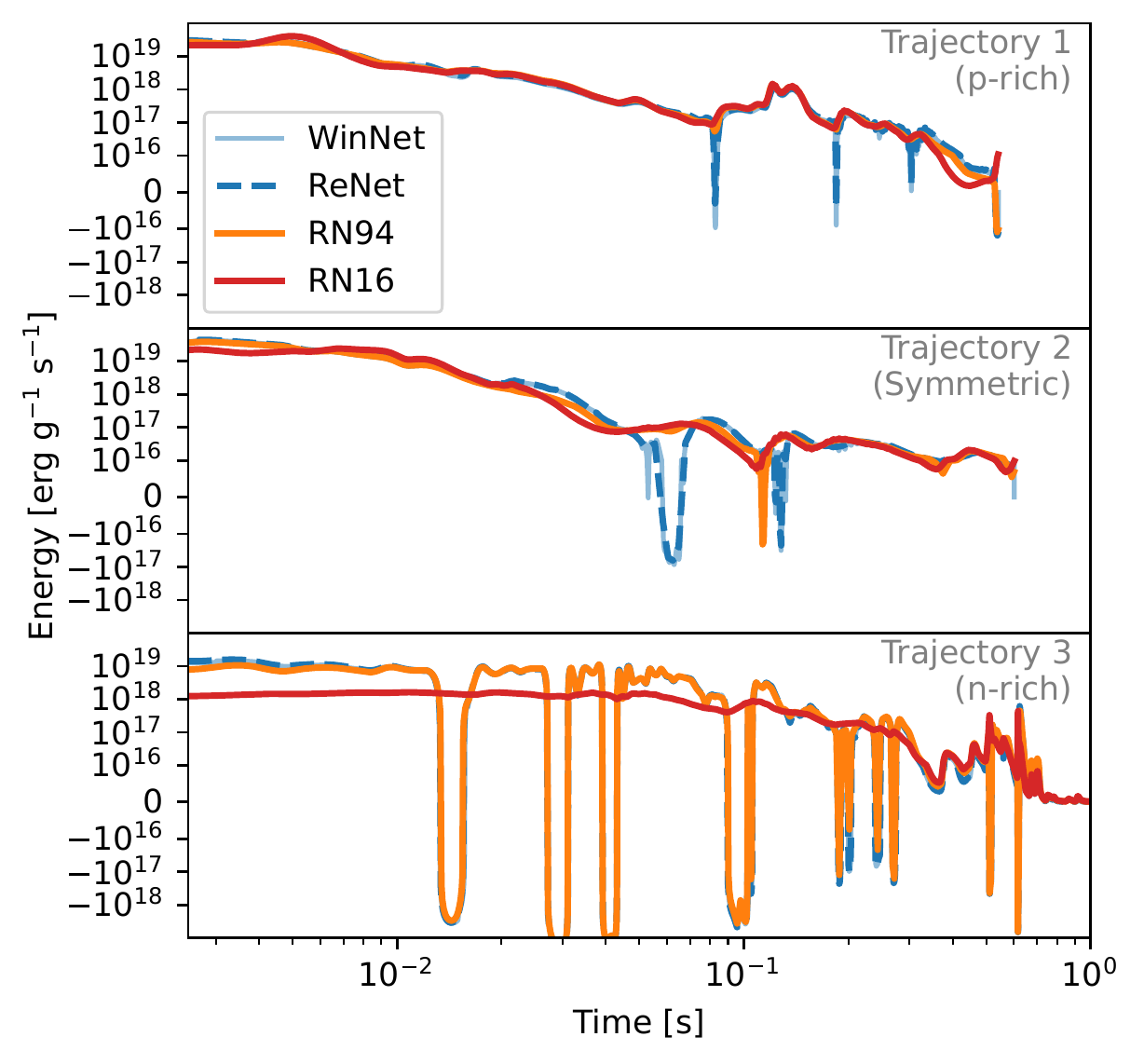}%
  \includegraphics[width=0.5\columnwidth]{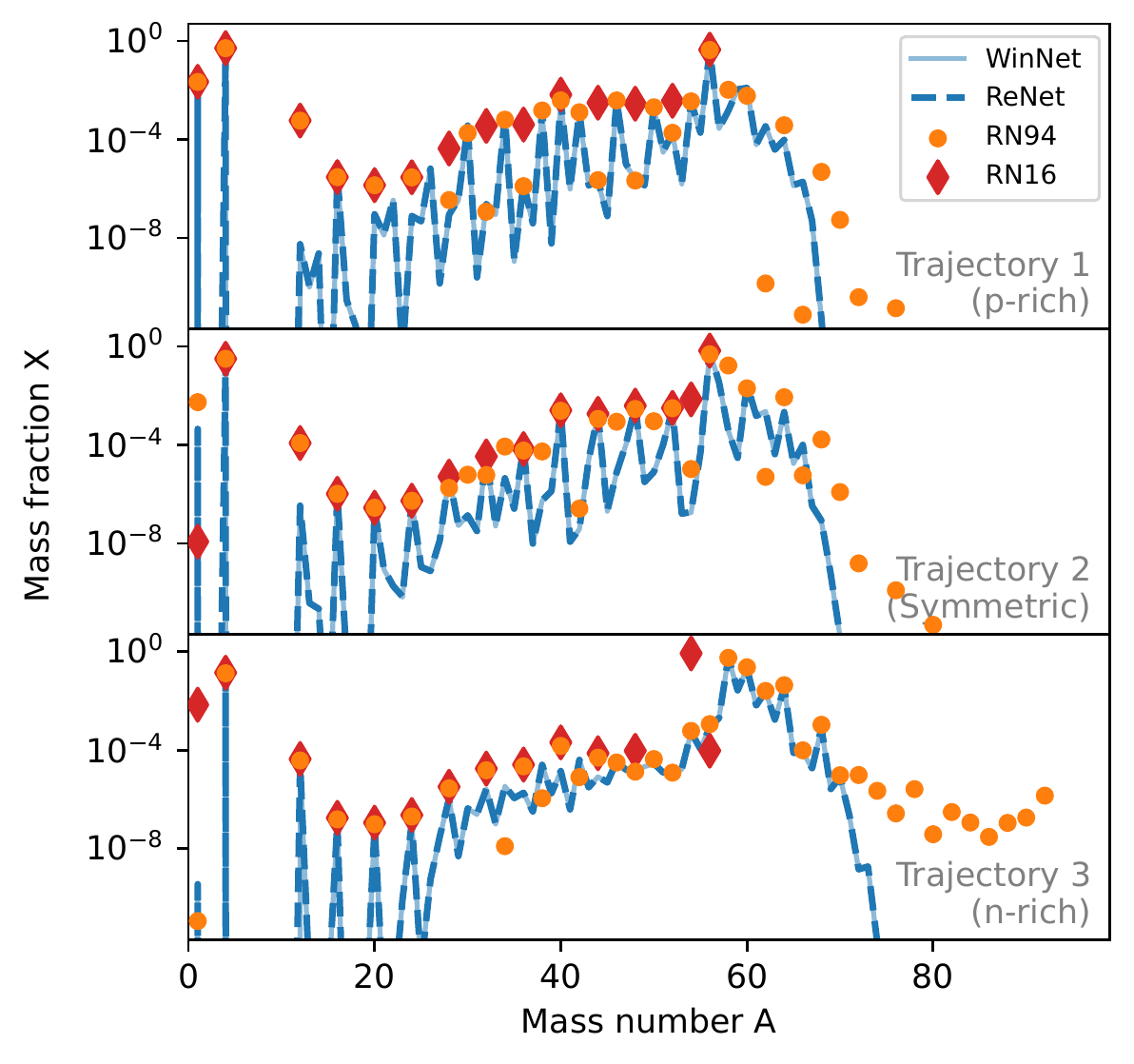}
    \caption{Left figure: Energy generation for two different trajectories. Shown is the generated nuclear energy for four different network architectures. The upper panel shows the energy generation for a slightly proton-rich trajectory ($Y_{e, 5.8\,\mathrm{GK}} \sim 0.51$), the middle panel a trajectory with symmetric conditions and the lower panel a slightly neutron-rich one ($Y_{e, 5.8\,\mathrm{GK}}\sim 0.48$). The time is relative to the start of the network calculation (i.e., $T=5.8\,\mathrm{GK}$). Right figure: Mass fractions versus mass numbers at the end of the simulation for three different trajectories and four different networks. The top panel shows the result for a slightly proton-rich trajectory ($Y_{e, 5.8\,\mathrm{GK}} \sim 0.51$), the middle panel a trajectory with symmetric conditions, and the lower panel the result for a slightly neutron-rich one ($Y_{e, 5.8\,\mathrm{GK}}\sim 0.48$).}
    \label{fig:renet_final_pattern}
    \label{fig:renet_engen}
\end{figure}

We created a network with a minimum amount of nuclei, but still covering the relevant region in the nuclear chart. Specifically, we focused on reproducing the following quantities:
\begin{itemize}
    \item The nuclear energy $\dot{E}$.
    \item Abundances of protons, neutrons, and alphas as well as other key nuclei such as $^{44}$Ti and $^{56}$Ni.
    \item Quantitative agreement of the abundance pattern at the end of the simulation.
\end{itemize}
In addition, we tested the reliability of \textsc{ReNet}. Therefore, we generated a network with $\sim 800$ species (in the following referred to as "full network"), including all relevant nuclei up to Ge and reactions, however, not including weak reactions\footnote{Because the nuclear flow evolves fairly close to stability for CC-SNe and the simulation times are usually less than the half life of the synthesized nuclei, this is a reasonable approximation. However, for longer simulation times or very neutron-rich material that evolves far from stability this is a point to improve.}. This generated network is compared to the well established reaction network \textsc{WinNet} \citep[][]{winteler_winnet}, which was already used in previous publications \citep[e.g.,][]{winteler_winnet,Korobkin2012,Martin2015,Eichler2019,Bliss2020,moritz_magneto,Ristic2022}. For this, we use three different typical explosive trajectories taken from the 2D\_flsh model (see Table~\ref{Tab:models}). One trajectory is slightly neutron-rich ($Y_{e, 5.8\,\mathrm{GK}} \sim 0.48$), one is symmetric ($Y_{e, 5.8\,\mathrm{GK}} \sim 0.50$), and one is slightly proton-rich ($Y_{e, 5.8\,\mathrm{GK}} \sim 0.51$).
For \textsc{WinNet} and \textsc{ReNet} both, the nuclear energy as well as the final abundances pattern are in excellent agreement (Fig.~\ref{fig:renet_engen}) as expected for an identical input. 
For a slightly proton rich trajectory (i.e., $Y_{e, 5.8\,\mathrm{GK}}\sim 0.51$), the energy generation of all reaction networks quantitatively agrees. There are three episodes in which the energy generation becomes significantly lower or even negative for a full network (upper panel of Fig~\ref{fig:renet_engen}). This is caused by an increase in temperature and, as as a consequence, a strong negative energy contribution of $(\gamma,\mathrm{p})$ reactions. Even though approximately implemented in RN94, the energy generation differs significantly in these episodes. The larger amount of considered nuclei of RN94 in comparison to RN16 does not improve the accuracy of the energy generation because the differences are caused by photodissociation of nuclei that are not included in RN94. The final mass fractions benefit from the larger number of included nuclei (upper panel of Fig.~\ref{fig:renet_final_pattern}). The differences between RN16 and RN94 for nuclei with $28<A<52$ are a direct consequence of additionally included nuclei such as $^{30}$Si or $^{30}$S, located on the proton- and neutron-rich side of the diagonal in the nuclear chart (see Figure~\ref{fig:insitu_vs_exsitu}). For the given trajectory, these nuclei are most abundant. Since they are not included in RN16, the nuclear flow is forced to the included symmetric nuclei. This causes $^{44}$Ti, but also other nuclei, to be overestimated by more than three magnitudes from the result of the full network calculation. Given the fraction of around $6\%$ of ejected matter with $Y_{e, 5.8\,\mathrm{GK}}\ge 0.51$ in the 2D\_flsh model, this can have a major effect on $^{44}$Ti abundances at the end of the simulation. On the other hand, $^{44}$Ti is in excellent agreement ($0.9\%$) between RN94 and the full network. In addition, the amount of $^{56}$Ni differs only slightly for both reduced networks from the full network ($1.5\%$ and $1.6\%$ for RN94 and RN16, respectively). Hereby, the difference is even smaller than the one between \textsc{ReNet} and \textsc{WinNet} ($3.5\%$) and is of numerical origin.

The energy generation of a trajectory with symmetric conditions ($Y_{e, 5.8\,\mathrm{GK}}=0.5$) evolves similarly to the previously discussed proton-rich case. Again, the episodes of negative nuclear energy generation in the full network are caused by $(\gamma,\mathrm{p})$ reactions on nuclei that are neither included in RN16 nor in RN94. In contrast to the slightly proton-rich trajectory, the flow evolves along symmetric nuclei, and the RN16 is therefore a better approximation compared to the previous, proton-rich case. The agreement with the full network of $^{44}$Ti is with $3\%$ even better for RN16 than the agreement of $35\%$ of RN94. Similarly, $^{56}$Ni is with $10\%$ deviation more precise than for RN94 with $22\%$ deviation. On the other hand, protons are underrepresented in the composition of RN16. The number of nucleons is important for the neutrino opacities. 

Even though only $6\%$ of the matter of the 2D\_flsh run is neutron-rich ($Y_{e, 5.8\mathrm{GK}}\le 0.49$) it is still desirable to reproduce the energy generation and composition within the reduced networks as a local energy generation may also be able to change the dynamics of the simulation. For these conditions, RN94 is able to reproduce all major features of the nuclear energy generation while RN16 differs by around one order of magnitude (lower panel of Fig.~\ref{fig:renet_engen}). Furthermore, RN16 is unable to follow the episodes of negative energy generation, because the nuclear flow is dominantly located in the iron region which is not covered by RN16. Furthermore, due to the mostly symmetric nuclei that are included in RN16, a lower electron fraction can only be achieved by an increase of the neutron abundance. 
\begin{figure}
\centering
 \includegraphics[width=0.5\columnwidth]{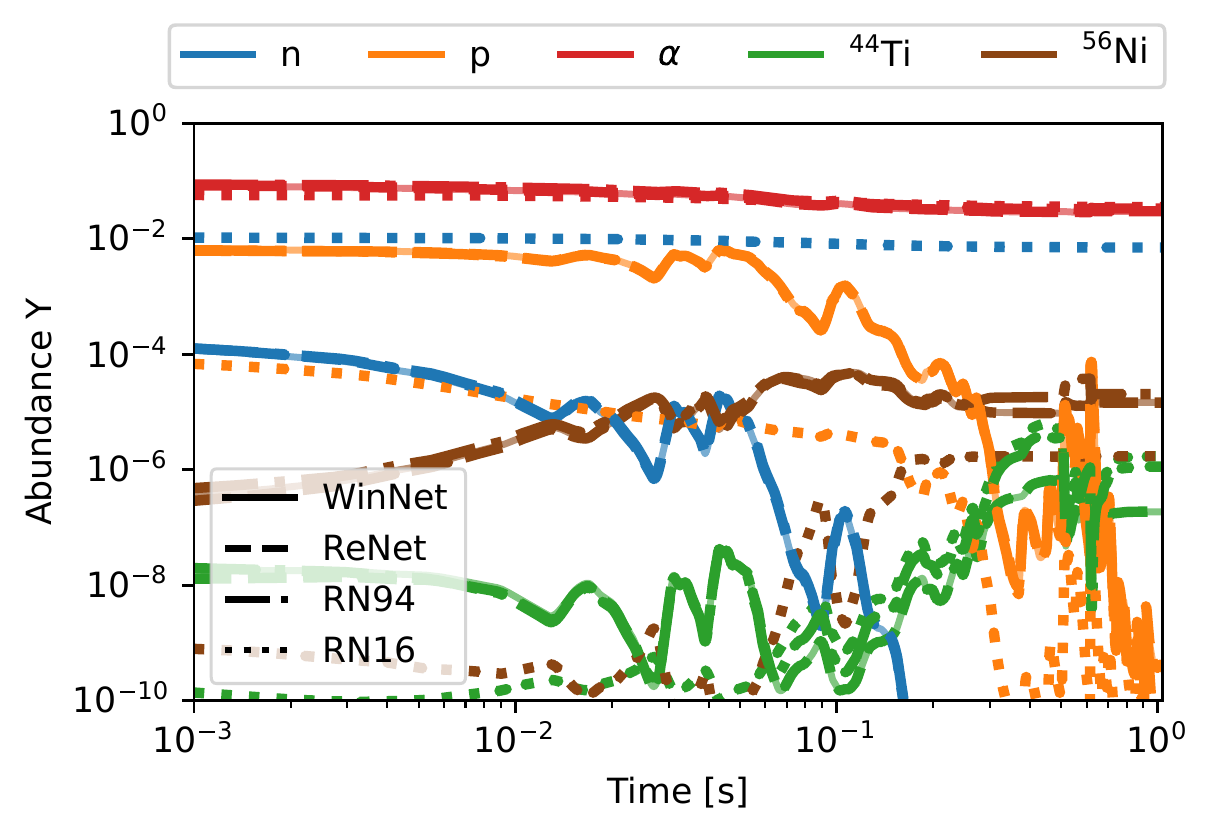}
 \caption{Evolution of neutrons, protons, alphas, $^{44}$Ti, and $^{56}$Ni for a neutron-rich trajectory. Shown are the results for four different reaction networks.}
 \label{fig:renet_npa}
\end{figure}
As a consequence, the NSE composition of neutrons and protons already differs significantly from that of the full network (Fig.~\ref{fig:renet_npa}). We note that this problem can be partly avoided by introducing another very neutron-rich nucleus (e.g., $^{56}$Cr within the aprox21, \citealt{Paxton2011}). The insufficient coverage of the nuclear chart of RN16 also leads to a larger deviation of $^{44}$Ti compared to RN94 (larger by a factor $8$ instead of a factor $5$) as well as a larger deviation of $^{56}$Ni ($88\%$ versus $41\%$). In general, the final abundances are better reproduced with RN94, especially the peak around the most abundant nucleus $^{60}$Ni (Fig.~\ref{fig:renet_final_pattern}). However, the reduced network synthesizes even heavier nuclei than the full reaction network. These heavy nuclei are even present when calculating the same trajectory, including the steady-state nuclei in the network. Therefore, this is not an effect of the steady state approximation, but rather of the exclusion of uneven nuclei in RN94. This exclusion of uneven nuclei in the calculation seems to bypass bottlenecks that are otherwise present and that prevent the flow from synthesizing heavier elements.

To summarize, RN16 and RN94 are good approximations to a full reaction network that includes all nuclei. For more extreme conditions toward neutron- and proton-rich conditions the larger reaction network RN94 provides a more accurate composition than the simpler RN16. The energy generation is well reproduced for both networks, with the exception of the neutron-rich conditions. 
\end{document}